\documentclass[cits]{PoS}
\usepackage{wrapfig}

\title{Recent Results of Cosmic Ray Measurements from IceCube and IceTop}

\ShortTitle{Recent Results of Cosmic Ray Measurements from IceCube and IceTop}

\author{
The IceCube Collaboration\footnote{For collaboration list, see PoS(ICRC2019) 1177.}\\
{\itshape \href{http://icecube.wisc.edu/collaboration/authors/icrc19_icecube}{http://icecube.wisc.edu/collaboration/authors/icrc19\_icecube}}\\
E-mail: \email{soldin@udel.edu}
}

\abstract{IceCube is a cubic-kilometer Cherenkov detector located in the deep ice at the geographic South Pole. The dominant event yield in the deep ice detector consists of penetrating atmospheric muons produced in cosmic ray air showers with energies above several $\mathrm{100\,GeV}$. In addition, the surface array, IceTop, deployed above the IceCube deep ice detector, measures the electromagnetic signal and low-energy muons of the air shower. Hence, IceCube and IceTop yield unique opportunities to study cosmic rays with large statistics in great detail.

We discuss the latest results of air shower measurements from IceCube and IceTop, including the energy spectrum of cosmic rays from $\mathrm{250\,TeV}$ up to the EeV range. We will also report a measurement of the cosmic ray mass composition in the PeV to EeV range and show recent results from searches for PeV gamma ray sources in the Southern Hemisphere. In addition, results from a full-sky analysis of the cosmic ray anisotropy, using combined data from the IceCube and HAWC observatories, will be reported. Finally, we will present a measurement of the density of muons in the GeV range and discuss its consistency with predictions from hadronic interaction models.\\

\vspace{4mm}
{\bfseries Corresponding author:}
\speaker{Dennis Soldin}$^{1}$\\
{$^{1}$ \itshape Bartol Research Institute, Dept. of Physics and Astronomy\\University of Delaware, Newark, DE 19716, USA}}

\FullConference{36th International Cosmic Ray Conference -ICRC2019-\\
		July 24th - August 1st, 2019\\
		Madison, WI, U.S.A.}

\begin{document}


\setcounter{page}{2}


\section{Introduction}
\label{sec:1}

High-energy cosmic rays enter the Earth's atmosphere with energies up to a few tens of EeV, producing extensive air showers that can be measured with large detectors on the surface and underground. The energy spectrum of cosmic rays and its features above a few $100\,\mathrm{TeV}$, such as the knee and ankle, are well known and have been measured with high statistical precision \cite{IceTop2,KASCADE1,Auger1,TA1,Tunka1}. However, several major questions in the understanding of cosmic rays remain open. Cosmic rays in the energy region from $\sim 100\,\mathrm{TeV}$ up to several $\sim 100\,\mathrm{PeV}$ are believed to be of galactic origin, but the sources of galactic cosmic rays remain unknown and their acceleration mechanisms are uncertain. The cosmic ray mass composition also remains uncertain \cite{KampertUnger}, especially towards the highest energies, where a transition from galactic to extra-galactic cosmic rays is expected \cite{H3a}. Although the highest energetic cosmic rays (above a few EeV) are believed to be of extra-galactic origin \cite{Auger2}, the sources and their acceleration mechanisms are also still unknown. In order to address these open questions, measurements of cosmic rays in combination with gamma ray and neutrino observations are of great importance where the observation of multiple astrophysical messengers can help to explain the acceleration mechanisms of high-energy cosmic rays and discover their sources.

\begin{figure}[b]
  \vspace{-0.3cm}
  
  \centering
  \mbox{\hspace{2. cm}\includegraphics[width=0.75\textwidth]{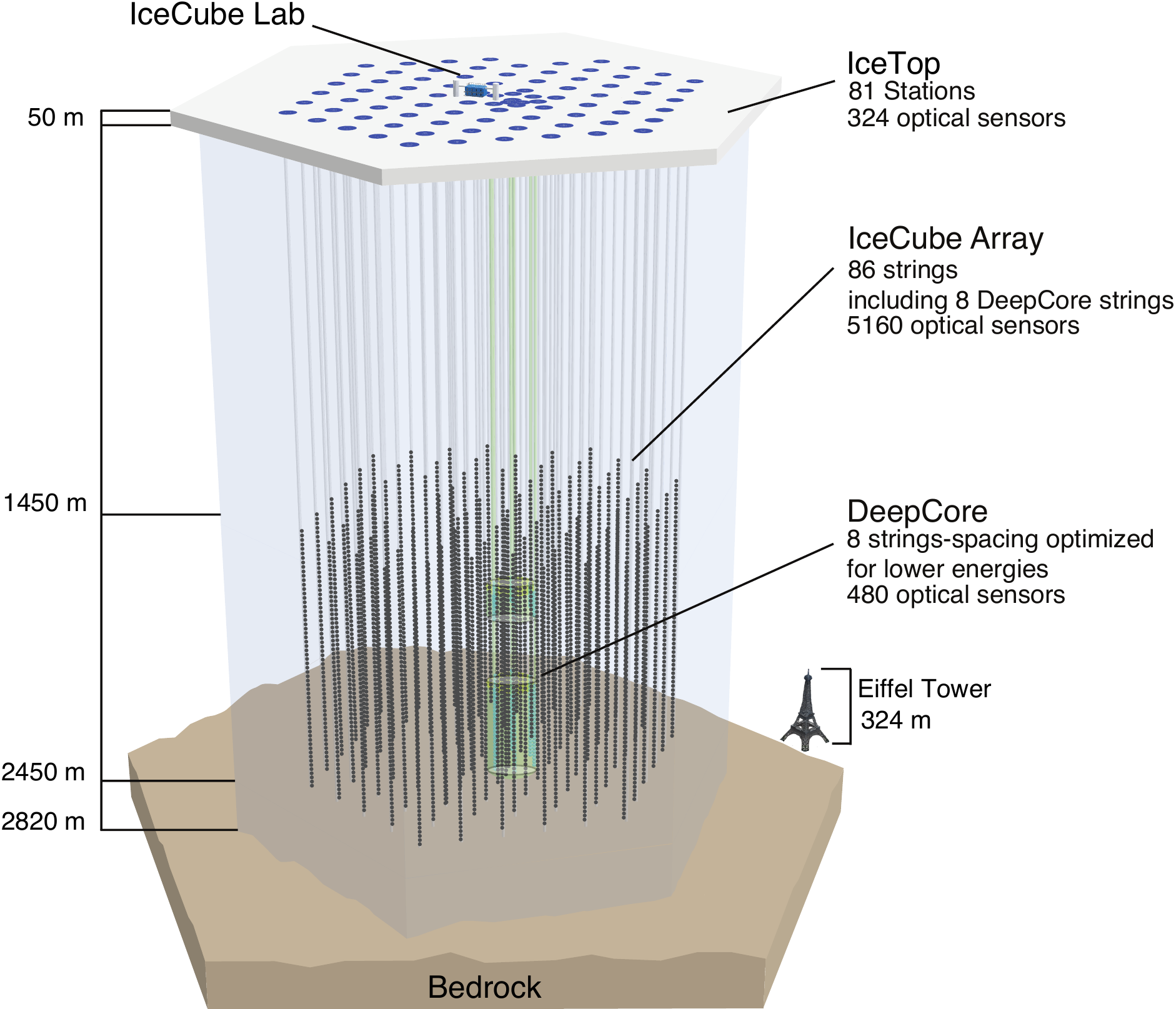}}
  \vspace{-0.1cm}
  
  \caption{Schematic drawing of the IceCube Neutrino Observatory \cite{IceCubeDet}. The deep ice detector consists of $5160$ DOMs \cite{IceCubeDet2} deployed on $86$ strings in depths of $1450\,\mathrm{m}$ to $2450\,\mathrm{m}$ in the Antarctic ice. The surface detector array IceTop comprises $81$ stations, consisting of two tanks with two DOMs each \cite{IceTopDet}.}
  \label{fig:ICArray}
  \vspace{-0.4cm}
  
\end{figure}

The IceCube Neutrino Observatory \cite{IceCubeDet} is located at the geographic South Pole and yields various opportunities to study cosmic rays in the context of multi-messenger astrophysics and beyond. It consists of a cubic-kilometer Cherenkov detector at depths of $1450\,\mathrm{m}$ to $2450\,\mathrm{m}$ in the Antarctic ice. As shown in Figure~\ref{fig:ICArray}, the deep ice detector comprises $86$ strings with $5160$ digital optical modules (DOMs) \cite{IceCubeDet2}. The strings are deployed in a hexagonal array with an average spacing of~$125\,\mathrm{m}$ and~$8$ strings are deployed with a denser average spacing of $\sim 72\,\mathrm{m}$, forming the subarray DeepCore. Each string hosts $60$ DOMs, containing a photomultiplier tube, calibration devices, and electronics for signal processing and readout. The average trigger rate of about~$2.15\,\mathrm{kHz}$ is mainly caused by high-energy atmospheric muons ($E_\mu\gtrsim 300\,\mathrm{GeV}$) which are generated in cosmic ray air showers in the Southern Hemisphere. The muons penetrate the Antarctic ice and produce Cherenkov light in the detector volume that is measured by the DOMs. Based on the light yield development in the ice, the initial direction of cosmic ray air showers, as well as the deposited energy of high-energy atmospheric muons, can be determined. 

The surface detector IceTop \cite{IceTopDet}, located about $\sim2.8\,\mathrm{km}$ above sea level (corresponding to an atmospheric depth of about $\sim~690\,\mathrm{g/cm}^2$) comprises $81$ stations, each consisting of two cylindrical Cherenkov tanks, which are separated by $10\,\mathrm{m}$, deployed approximately consistent with the location of the IceCube strings. Each tank is filled with clear ice and houses two DOMs (high- and low-gain) that measure the Cherenkov light generated by charged, relativistic particles from cosmic ray air showers when they traverse the tank. An infill area in the center of the detector has a denser spacing of $<50\,\mathrm{m}$, that is used to improve the sensitivity of air shower detection towards  low energies. While the deep ice detector measures the high-energy muons from air showers, IceTop is primarily sensitive to the electromagnetic component and muons with energies around a few~GeV. This information can be used to determine the initial direction and total energy of the air showers. Thus, the complementary information from both detectors yields unique opportunities to study cosmic ray air showers in the energy range between a few hundred TeV up to the EeV range in great detail.

This paper discusses the recent results from measurements of cosmic rays with IceCube and IceTop. This includes measurements of the spectrum and composition, as well as the muon content of cosmic ray air showers. Moreover, an analysis of the arrival directions of cosmic rays and searches for high-energy gamma rays from the Southern Hemisphere will be presented. An overview of analyses of high-energy atmospheric muons in the deep ice detector can be found in Ref.~\cite{ISVHECRIDennis}. The latest results from various other measurements with IceCube are reported separately in Ref.~\cite{ICRCDawn}.



\section{Cosmic Ray Energy Spectrum}
\label{sec:2}

The energy spectrum of cosmic rays can be determined from the air shower signals measured by the IceTop tanks. For the analysis of IceTop data, only events that pass a cosmic ray filter are studied, and all signals are calibrated accounting for the specific tank responses, as described in detail in Ref.~\cite{IceTopDet}. Following calibration, all IceTop signals are expressed in units of expected vertical equivalent muons (VEM) and various basic event cleanings are applied. 


A dedicated air shower reconstruction is applied to the data \cite{IceTopDet} where the core position and direction of the air shower are determined and its lateral profile is fit by a lateral distribution function (LDF) of the form
\begin{equation}
S(r)=S_{125}\cdot\left(\frac{r}{125\,\mathrm{m}}\right)^{-\beta-\kappa\cdot \log_{10}(r/125\,\mathrm{m})}\, .
\label{eq:S125}
\end{equation}
This function describes the charge distribution of the event, $S(r)$, in units of VEM as a function of the distance to the shower axis, $r$. The free parameter, $\beta$, measures the steepness of the LDF and $S_{125}$ fits the signal strength at a reference distance of $125\,\mathrm{m}$ from the shower axis. The parameter $\kappa$ is a measure of the curvature of the shower front, which is assumed to be approximately constant with $\kappa=0.303$. The corresponding timing distribution is described by a paraboloid with a gaussian nose, as defined in Ref.~\cite{IceTop2}.


Due to the environmental conditions at the South Pole, IceTop tanks are buried under a layer of snow that needs to be accounted for during air shower reconstruction. The depth of the snow depends on the tank location and the time of data taking and varies between several tens of centimeters and a few meters (see Ref.~\cite{IceTopDet} for details). A simple exponential reduction of each measured signal is applied to the expected signal strength $S(r)$, as described in Ref.~\cite{IceTop2}. The uncertainties due to this procedure are well studied and included in the detector systematics.



The best fit LDF parameters $\beta$ and $S_{125}$ are found using a three-step maximum-likelihood technique, comparing charges and timing of the cleaned hits of each event to the expected distributions for the charges (Equation (\ref{eq:S125})) and timing, as described in Ref.~\cite{IceTopDet}. After a successful reconstruction (where a maximum likelihood value is found) further quality cuts are applied, which are described in detail in Refs.~\cite{IceTop2,IceTop3}. In addition, the zenith angle distribution of surviving events is constrained to angles below $\sim 37^\circ$. In order to get an estimate of the initial cosmic ray energy, the energy proxy $S_{125}$ (in units of VEM) is converted into shower energy $E_0$ (in units of GeV). This is done via CORSIKA simulations \cite{CORSIKA} with Sibyll~2.1 \cite{Sib21} as the hadronic interaction model and with an H3a primary flux assumption \cite{H3a}. 

As described in Ref.\cite{IceTop3}, the relation between shower energy and $S_{125}$ can be parametrized with a double logarithmic linear function for different zenith angle regions. These functions allow for an estimation of the shower energy for each event. Although the detector efficiency reaches~$100\%$ for all primary nuclei at energies near $\sim 3\,\mathrm{PeV}$, the reconstruction efficiency needs to be taken into account as high-energy showers can exceed the size of the IceTop footprint and the reconstruction becomes less precise. This is done by correcting for the efficiencies obtained from the simulated events, as described in Ref.~\cite{IceTop2}. Using these techniques, the energy resolution in this analysis is below $0.1$ in $\log_{10}(E_0/\mathrm{GeV})$, and the angular resolution is better than $\sim 1^\circ$.

\begin{figure}[b]
\vspace{-0.1cm}

  \mbox{\hspace{-0.3 cm}
  \includegraphics[width=.49\textwidth]{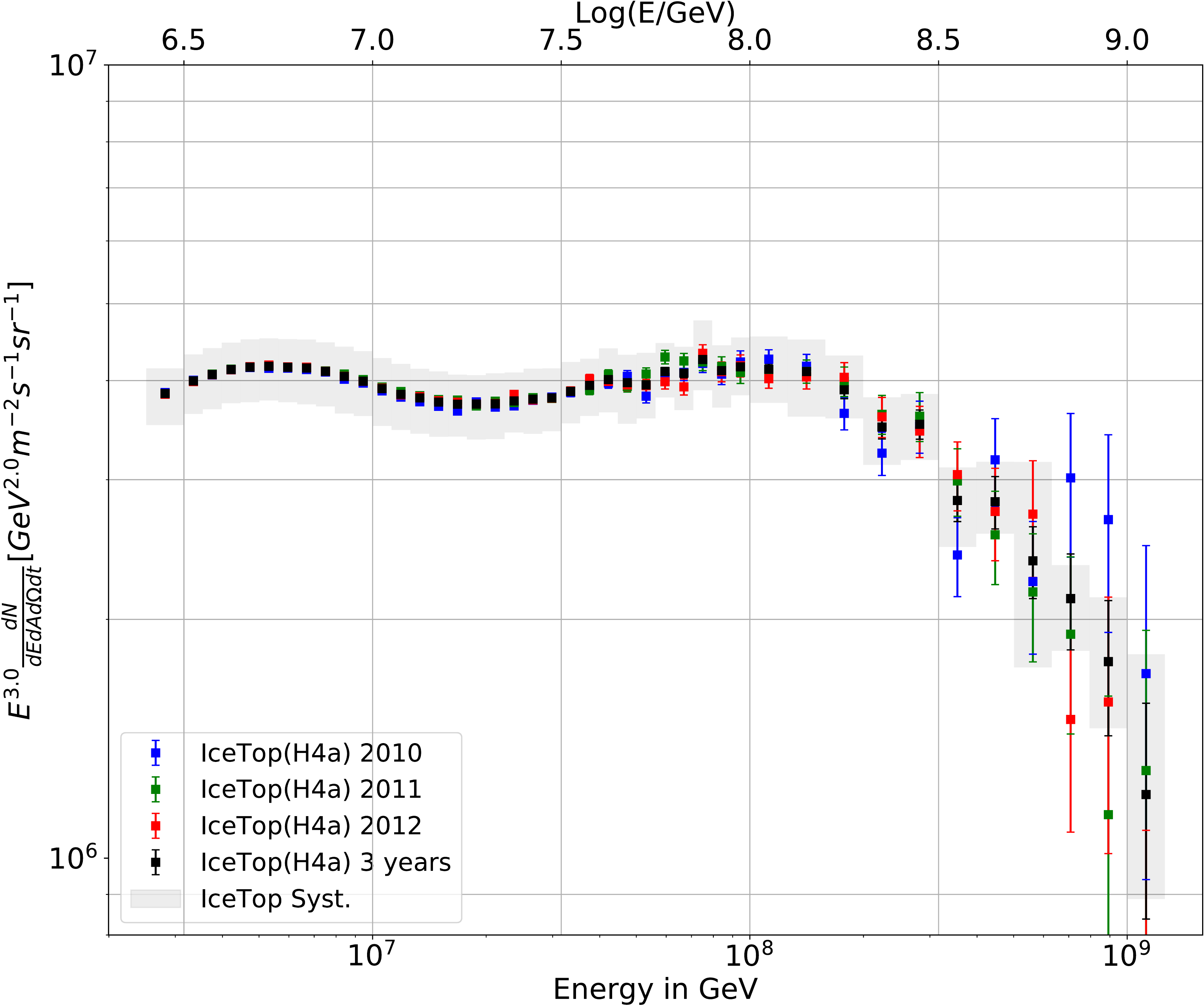}
  \hspace{0.1 cm}
  \mbox{
  \includegraphics[width=.503\textwidth]{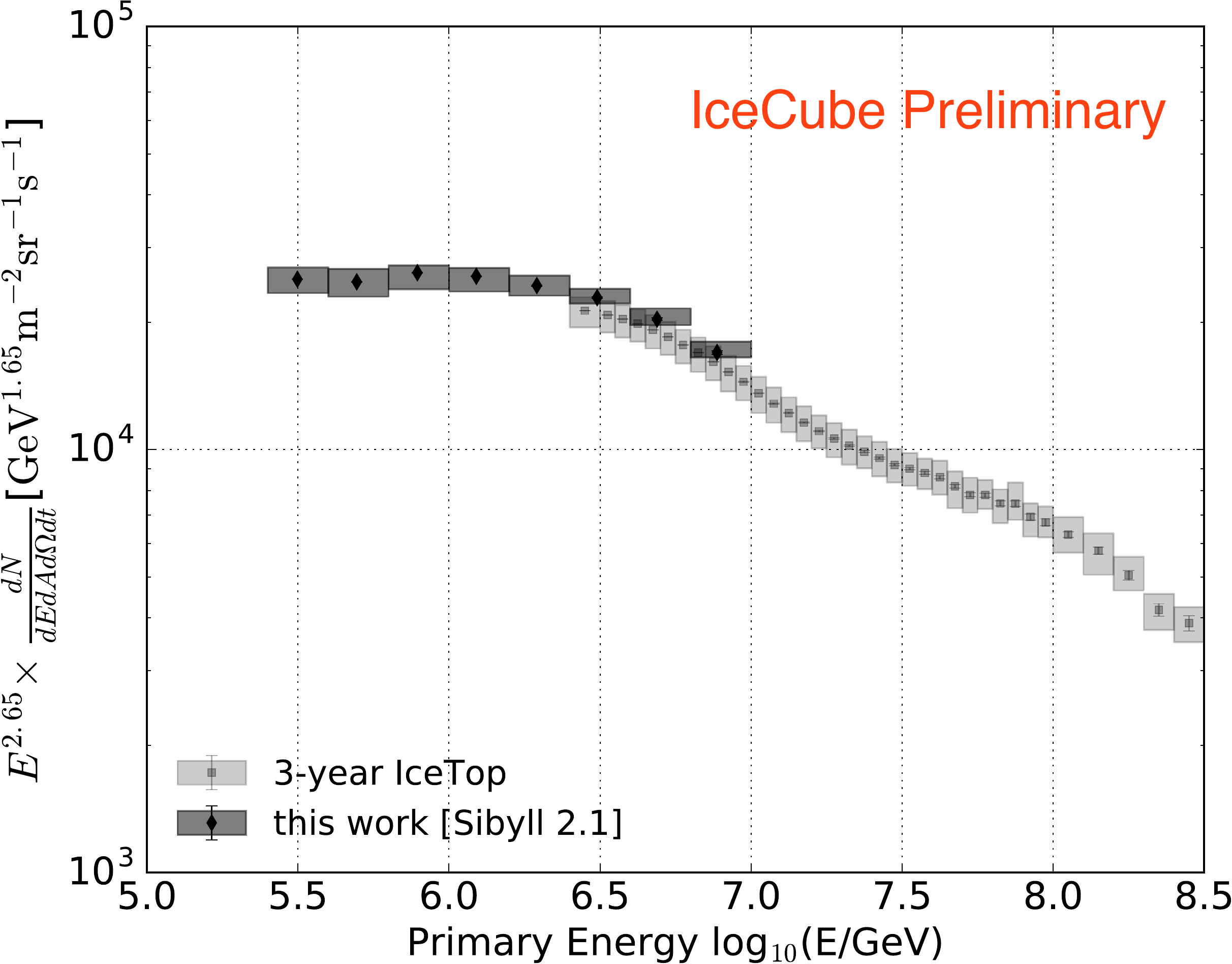}\vspace{0.2cm}}}
  \vspace{-0.5cm}
  
  \caption{Left: Cosmic ray spectrum obtained from three years of IceTop data \cite{IceTop3}. The grey band indicates the corresponding systematic uncertainties. Right: Low-energy extension of the cosmic ray spectrum obtained using random forest regression methods \cite{IceTop5}, as described in the text.}
  \label{fig:CRSpectrum}
  \vspace{-0.4cm}
  
\end{figure}

A measurement of the cosmic ray spectrum was performed using IceTop data from June 2010 through May 2013, about $\sim 5.1\cdot 10^7$ surviving events \cite{IceTop3}.
The resulting spectrum is shown in Figure~\ref{fig:CRSpectrum} (left) and it agrees well with previous measurements within the uncertainties \cite{IceTop2}. Systematic detector uncertainties are shown as a grey band and include the VEM calibration and snow depth uncertainties, as discussed in Ref.~\cite{IceTopDet}.

In order to extend the energy spectrum towards lower energies, a dedicated event selection is necessary. This event selection is based on the denser infill area of IceTop, as described in detail in Ref.~\cite{IceTop5}, that lowers the energy threshold for air shower events in IceTop by roughly one order of magnitude. The reconstruction methods described above, however, are not feasible for events with only a few hit stations. The reconstruction of low-energy events is, instead, based on iterative random forest regression techniques \cite{scikit-learn}, as described in detail in Ref.~\cite{IceTop5}. The random forest is trained using $50\%$ of the simulated events and the other $50\%$ for testing and performance optimization which is done using a cross-validation grid search. The CORSIKA simulations used during the training and testing procedure use Sibyll~2.1 as a hadronic model and they assume an H3a primary cosmic ray flux. In order to account for the efficiencies of this analysis, effective areas are used that are obtained from these simulations. In addition, a Bayesian iterative unfolding~\cite{pyUnfold} is applied to the data to account for potential systematic  effects of this analysis ({\it e.g.} bin migration). Since the random forest is trained under certain model assumptions, systematic biases from the flux model, the atmospheric model, as well as from the effective area calculation and the unfolding procedure are considered as systematic uncertainties. The spectrum is derived separately for different hadronic interaction model assumptions, and results based on QGSJet II-04 \cite{QGSJet} as the hadronic model are discussed in Ref.~\cite{IceTop5}.

The low-energy spectrum, using data from May 2016 to May 2017 ($7.4\cdot 10^6$ events) \cite{IceTop5}, is shown in Figure~\ref{fig:CRSpectrum} (right) in comparison to the three-year result discussed above. Both results agree in the overlap region of $6.5\leq \log_{10}(E_0/\mathrm{GeV})\leq 7.0$ within their uncertainties. Using this technique, the cosmic ray energy spectrum measured by IceTop is extended down to $\sim 250\,\mathrm{TeV}$ with an energy resolution better than $0.2$ in $\log_{10}(E_0/\mathrm{GeV})$.


\section{Cosmic Ray Mass Composition}
\label{sec:3}

The muon content of air showers is sensitive to the mass of the initial cosmic ray. According to the Matthews-Heitler model \cite{Heitler}, the muon number, for example, scales with the mass number, $A$, and cosmic ray energy, $E_0$, approximately as
\begin{equation}
N_{\mu}(E_0,A)\propto A\cdot \left(\frac{E_0}{A}\right)^\beta=A^{1-\beta}\cdot E_0^\beta\, ,
\label{eq:NMu_A}
\end{equation}
with $\beta\simeq 0.9$. In addition, protons are more likely to produce extremely high-energy muons than heavier primary nuclei, that can have large local energy deposits due to radiative processes, such as Bremsstrahlung and pair-production \cite{IceCube1}. Hence, dedicated measurements of the muon content of air showers via their stochastic energy losses in the deep ice detector, together with the IceTop information, can be used to obtain an estimate of the mass spectrum of cosmic rays.

\begin{figure}[t]
\vspace{-0.9 cm}

  \mbox{\hspace{-0.4 cm}
  \includegraphics[width=0.64\textwidth]{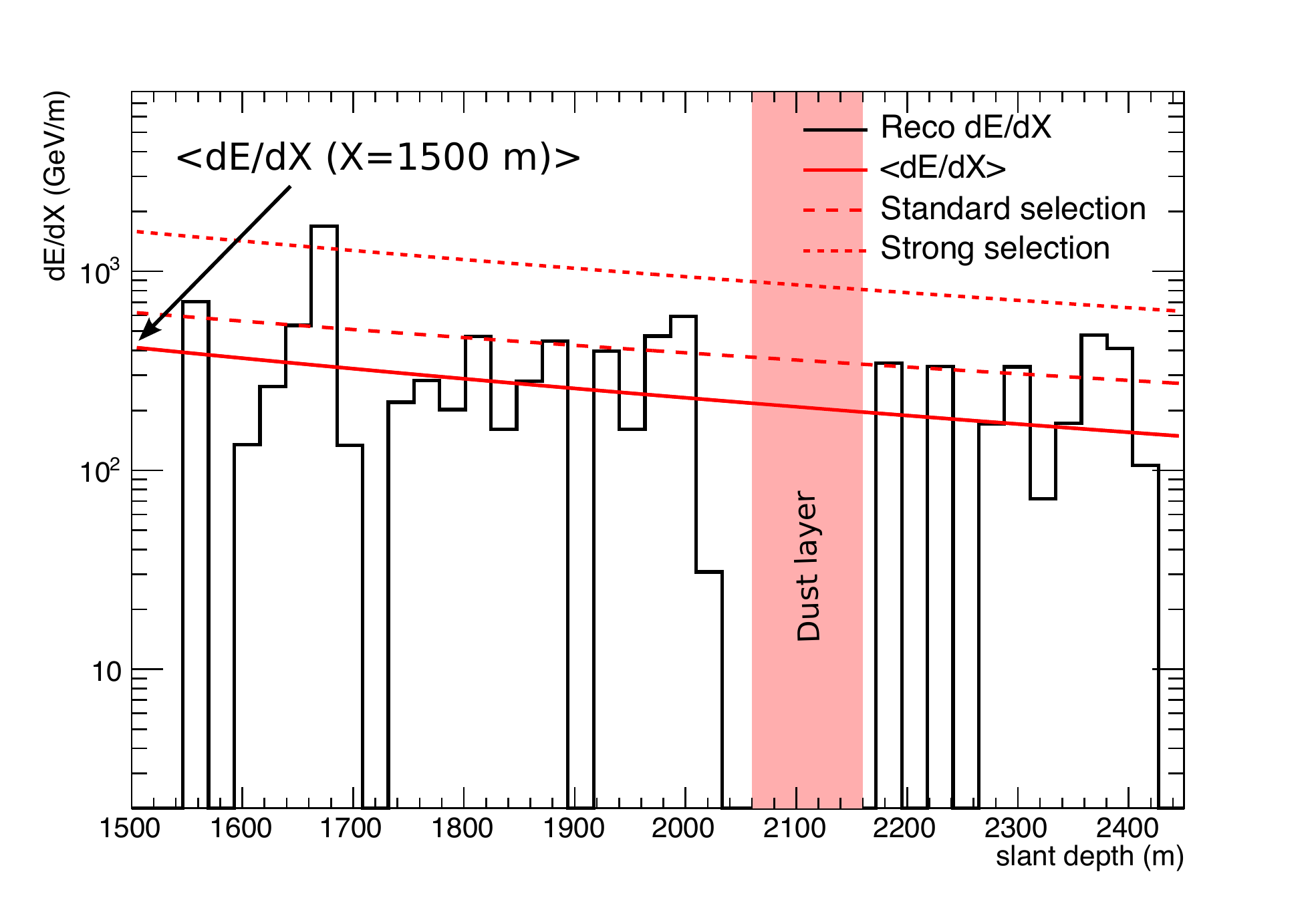}\hspace{-0.6cm}
  \includegraphics[width=0.442\textwidth]{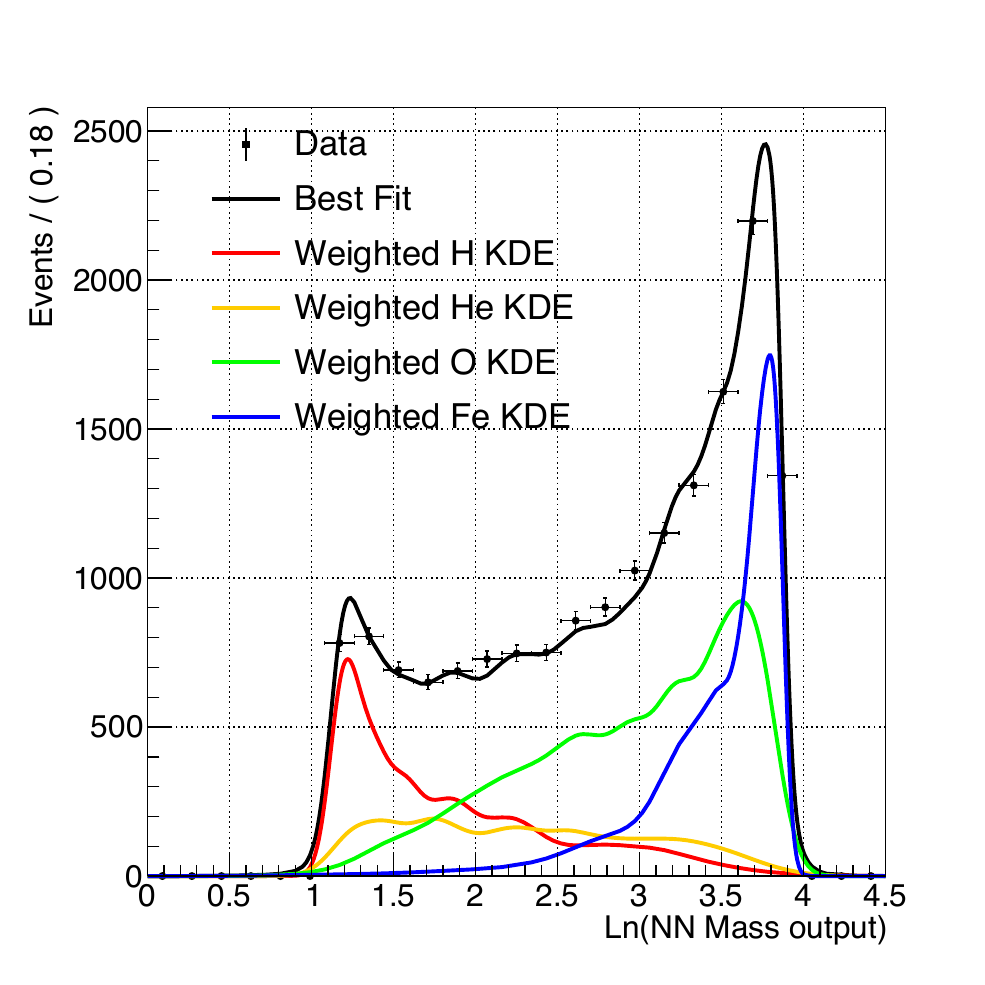}}
  \vspace{-0.7cm}
  
  \caption{Left: Energy loss profile obtained using the reconstruction algorithm described in Ref.~\cite{Millipede}. In addition to the fit to the continuous energy loss (solid curve), two different thresholds used to count large stochastic losses are also shown (dashed and dotted curves), as described in the text. Right: Mass output from the neural network for the energy bin $7.4\leq \log_{10}(E_0/\mathrm{GeV})\leq 7.5$. Colored lines show the PDF templates that are fit to the data in order to obtain the mass composition (see text for details) \cite{IceTop3}.}
  \label{fig:CRMassTemplate}
  \vspace{-0.3cm}
  
\end{figure}

Such an analysis was performed using coincident events, observed by IceTop and IceCube~\cite{IceTop3}. All events are kept that pass any IceTop cosmic ray filter \cite{IceTopDet}, and have more than $8$ hits recorded in the deep ice detector. The standard selection criteria are applied to the IceTop data, as described for the analysis of the cosmic ray spectrum spectrum above. Only events with a successful IceTop reconstruction are kept, and the deposited energy along the extrapolated IceTop trajectory is derived based on the charge and timing information in the ice. This is done using a dedicated reconstruction algorithm that is described in detail in Ref.~\cite{Millipede}. An example of a reconstructed energy loss profile is shown in Figure~\ref{fig:CRMassTemplate} (left). The resulting energy loss profile of each event can be characterized by its average energy loss behavior along the track and deviations from this average behavior. The reconstructed energy loss at a reference slant depth of $1500\,\mathrm{m}$ in the ice, as well as two measures of the number of high-energy stochastic losses along the track, as defined in Ref.~\cite{IceTop2}, are used as mass sensitive variables for further analysis. 

These observables, together with the energy proxy $S_{125}$ and the reconstructed zenith angle from IceTop, are used as input for an artificial neutral network \cite{TMVA} to determine the relationships between the five input features and the two output parameters (shower energy and mass estimate). The neural network is trained using $50\%$ of the simulated events while the other $50\%$ is used to test the network and optimize its performance. The underlying simulated dataset was generated with CORSIKA using Sibyll~2.1 and it contains an equal fraction of H, He, O, and Fe primary cosmic rays. The neural network was optimized by using various network architectures with different input variables and with a number of layers and neurons, as described in Ref.~\cite{IceTop3}. The output parameters are continuous and the mass output is distributed around the four discrete mass numbers. The distributions are converted into template probability density functions (PDFs) using an adaptive kernel density estimation (KDE) method \cite{KDE} and are used to fit the data separately in each energy bin. An example PDF fit is shown in Figure~\ref{fig:CRMassTemplate} (right) for the energy bin $7.4\leq \log_{10}(E_0/\mathrm{GeV})\leq 7.5$.

 \begin{figure}[t]
 \vspace{-0.5cm}
 
  \mbox{\hspace{-0.36 cm}
  \includegraphics[width=0.51\textwidth]{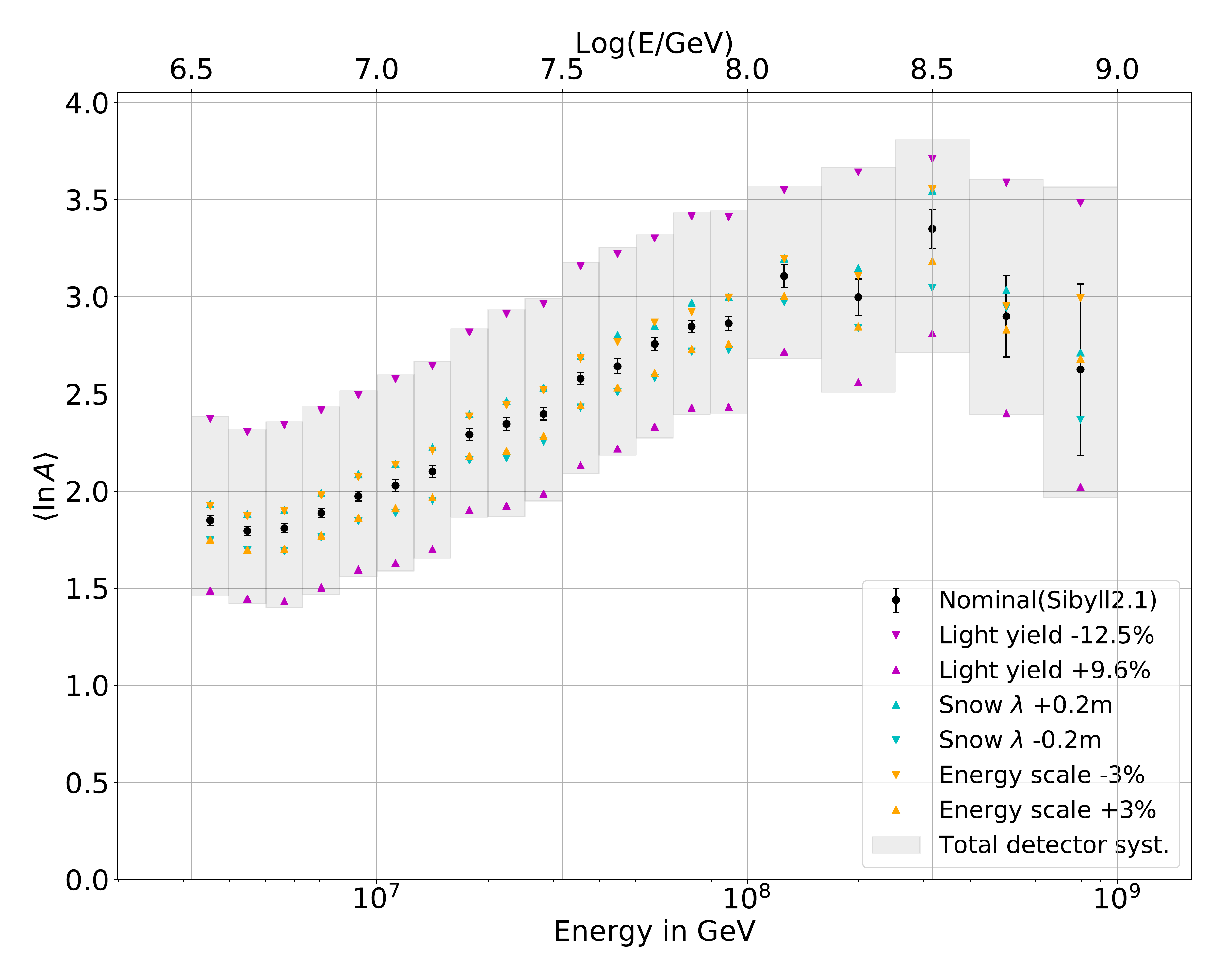}
  \includegraphics[width=0.51\textwidth]{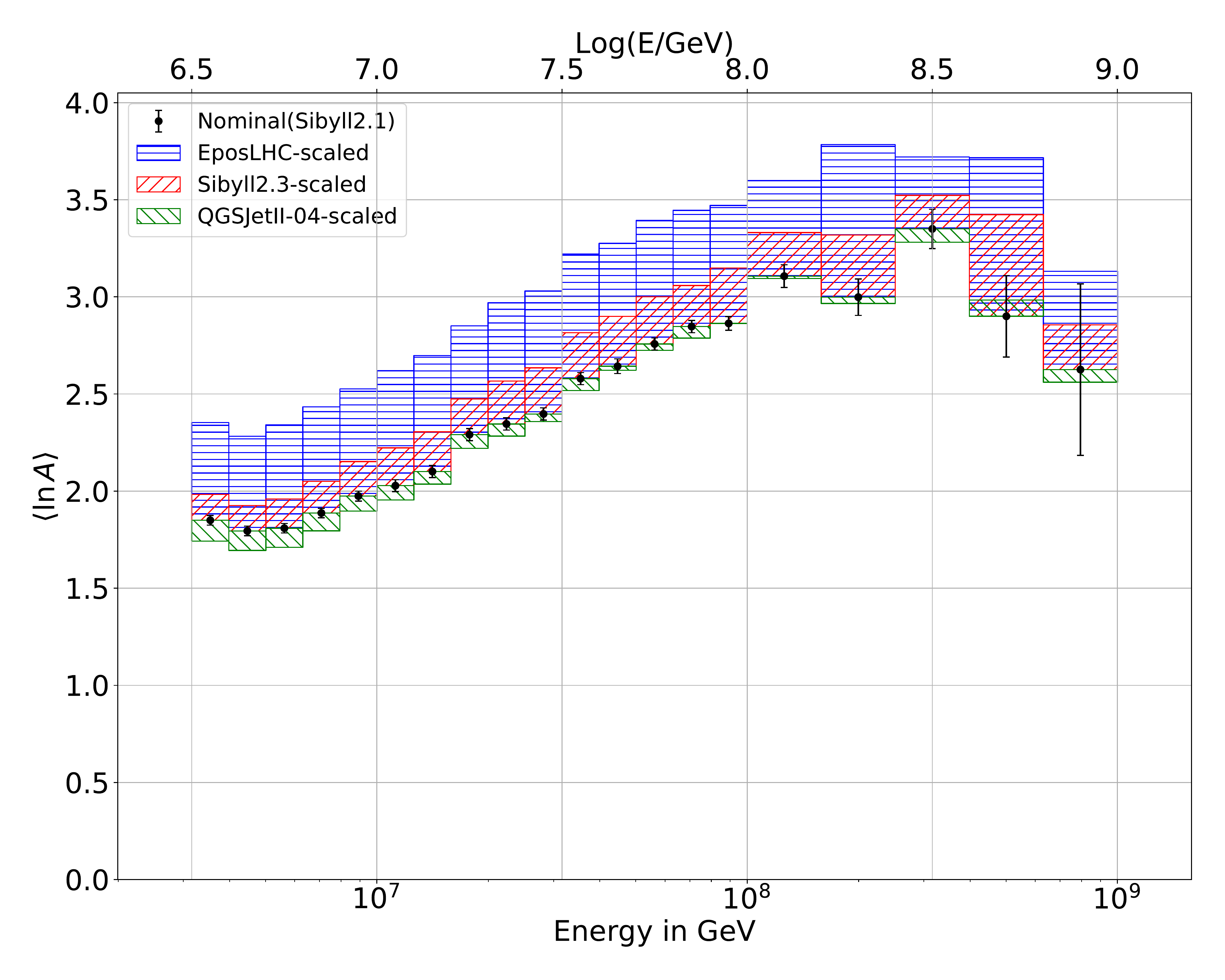}}
  \vspace{-0.6cm}
  
  \caption{Mean logarithmic mass of cosmic rays obtained from three years of IceTop and IceCube data \cite{IceTop3}. Several detector uncertainties are shown separately (right) and an estimate on the uncertainties from hadronic interaction models is also shown (right).}
  \label{fig:CRMass1}  
\end{figure}

The resulting mean logarithmic mass spectrum, obtained from data taken by IceCue and IceTop between June 2010 and May 2013 ($\sim 7.3\cdot 10^6$ events), is shown in Figure~\ref{fig:CRMass1} \cite{IceTop3}. Also shown are the systematic detector uncertainties (left), including effects of the snow accumulation, the absolute energy scale of IceTop, and the light yield measured in the ice. Uncertainties due to the hadronic interaction model are shown separately in Figure~\ref{fig:CRMass1} (right). These are obtained from statistically limited CORSIKA simulations using Sibyll 2.3 \cite{Sib23}, QGSJet II-04, and EPOS-LHC \cite{EPOS} as hadronic models. These simulations are generated with only $10\%$ of the statistics with respect to the baseline Sibyll~2.1 dataset where only proton and iron showers are produced due to the computational efforts to produce the datasets. This makes it challenging to repeat the full analysis with these simulations and the effect of hadronic models is instead estimated utilizing the differences observed in $S_{125}$ and the deposited energy in the ice. This method derives scaling factors for various hadronic models that are applied to the experimental data, and the full analysis is applied to this dataset. The result is then weighted according to the elemental mass fractions obtained from Sibyll~2.1 simulations, as described in Ref.~\cite{IceTop3}.

\begin{figure}[t]
\vspace{-0.8cm}

  \mbox{\hspace{0.1 cm}\includegraphics[width=0.98\textwidth]{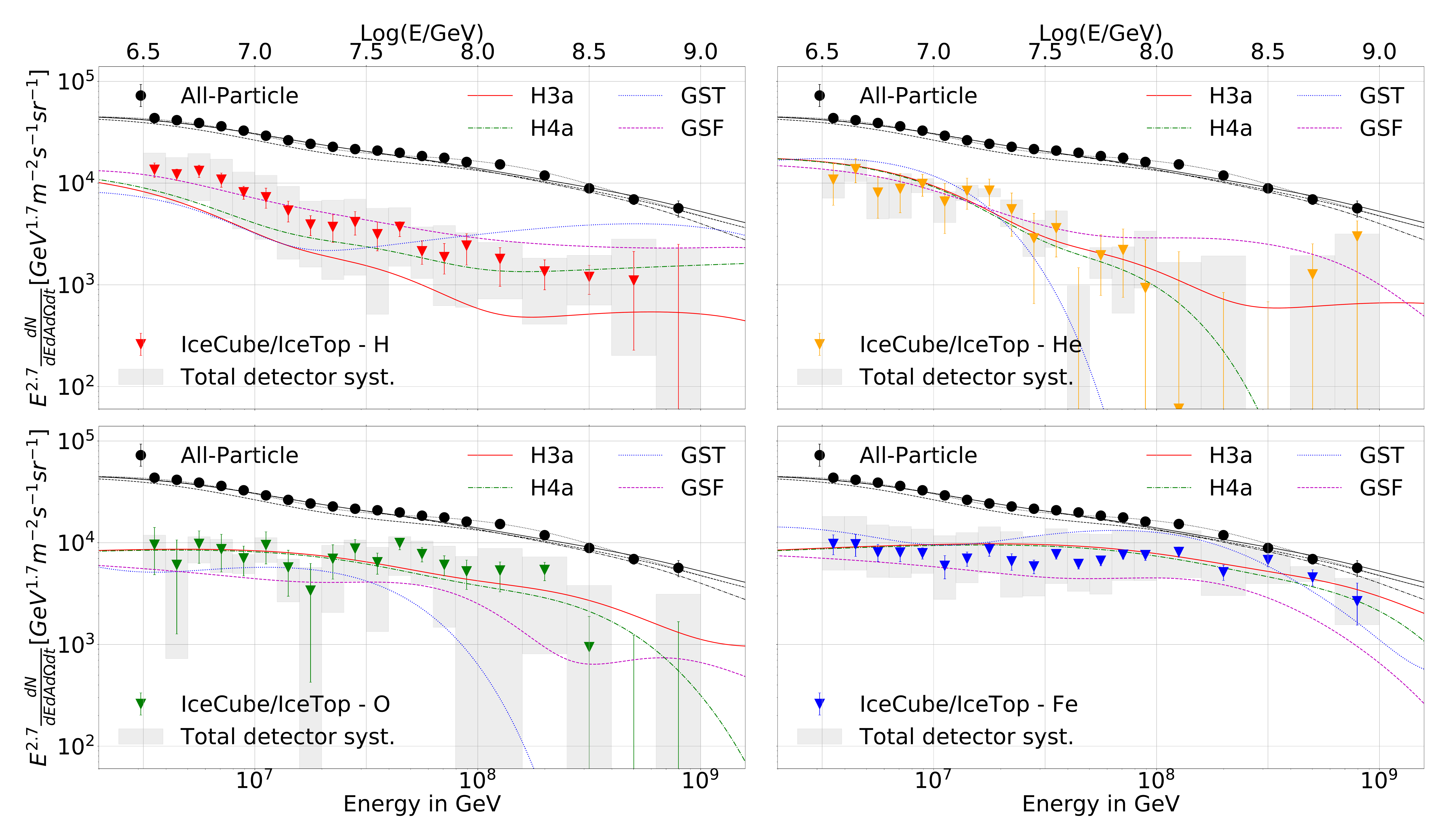}}
  \vspace{-0.5cm}
  
  \caption{Elementary mass spectra obtained from three years of data from IceCube and IceTop for the four mass groups considered in this analysis \cite{IceTop3}. Also shown are the systematic uncertainties described in the text and model predictions from Refs.~\cite{H3a,GST,GSF}.}
  \label{fig:CRMass2}
  \vspace{-0.5cm}
  
\end{figure}

The resulting cosmic ray all-particle spectrum, as well as the individual mass spectra for the four mass groups used in this analysis, are shown in Figure~\ref{fig:CRMass2}. As discussed in Ref.~\cite{IceTop3}, the all-particle spectrum from this analysis agrees well with the IceTop-alone spectrum, described in Section \ref{sec:2}, which is obtained using independent analysis methods. As shown in Figure~\ref{fig:CRMass2}, different flux models show tension towards high energies and this measurement will help to improve future models based on empirical data.

\section{Searches for PeV Gamma Ray Sources}
\label{sec:4}

High-energy cosmic rays escape their local source environment and interact with interstellar gas while propagating through the galaxy, producing high-energy gamma ray emission. This emission is expected to be concentrated along the galactic plane where most of the nearby interstellar gas is located. Measurements of this diffuse emission can help to explain cosmic ray diffusion processes in the galaxy \cite{Kalberla1}. In the energy region of $\sim 600\,\mathrm{TeV}$ to about $100\,\mathrm{PeV}$, gamma rays can only reach the Earth from length scales comparable to the size of the galaxy due to the large cross-section for pair production with the cosmic microwave background (CMB) \cite{Biermann1}. In this energy regime, IceCube and IceTop are sensitive to the detection of gamma rays \cite{IceCube2}, and are the sole experiment to be sensitive to PeV gamma rays in the Southern Hemisphere.

Similar to the cosmic ray mass composition analysis, as described in Section~\ref{sec:3}, the main discriminator to select gamma ray events is the number of muons produced in an air shower. While air showers initiated by charged cosmic rays produce muons predominantly in hadronic processes, in gamma ray air showers they are primarily produced through pair production and the decay of photo-produced pions and kaons \cite{Halzen1}. These processes are, however, highly suppressed and therefore gamma ray induced air showers are typically characterized by their low muon content. 

All IceTop events that pass a cosmic ray filter are selected for further analysis of gamma ray events, and basic events cleanings are applied to the data using the same techniques as described in Section~\ref{sec:2}. The standard IceTop hit selection removes single hit tanks and since these hits are predominantly produced by muons, especially towards large distances from the shower axis (see also Section~\ref{sec:6}), this information is utilized in this analysis. In addition, information from the deep ice detector is also used for this analysis if more than $8$ hits have been recorded. Subsequently, basic hit cleanings (described in Ref.~\cite{IceCube3}) are applied in order to reduce the number of uncorrelated hits due to noise.

\begin{figure}[t]
\vspace{-0.3cm}

  \mbox{\hspace{-0.3cm}\includegraphics[width=0.53\textwidth]{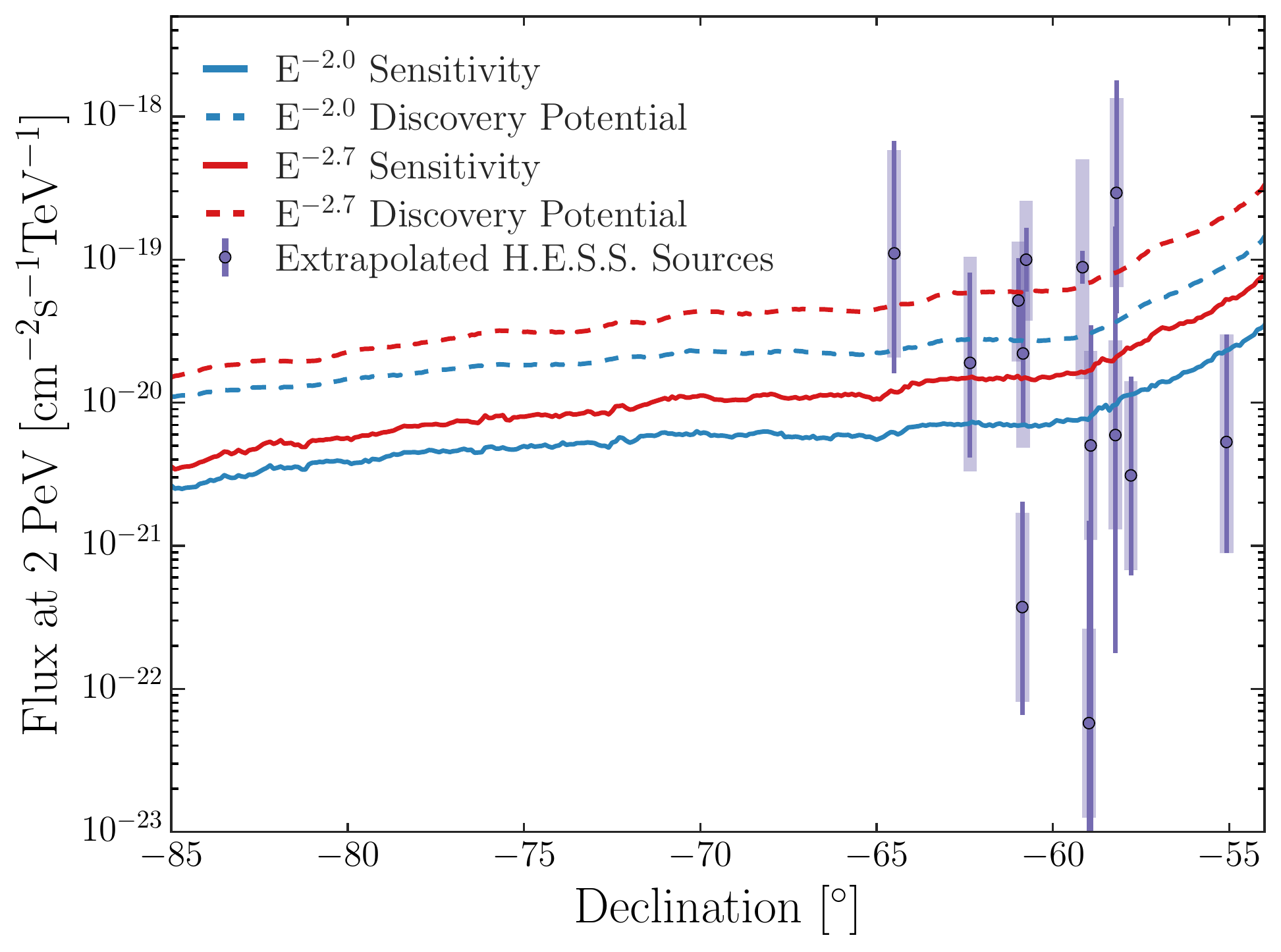}
  \includegraphics[width=0.471\textwidth]{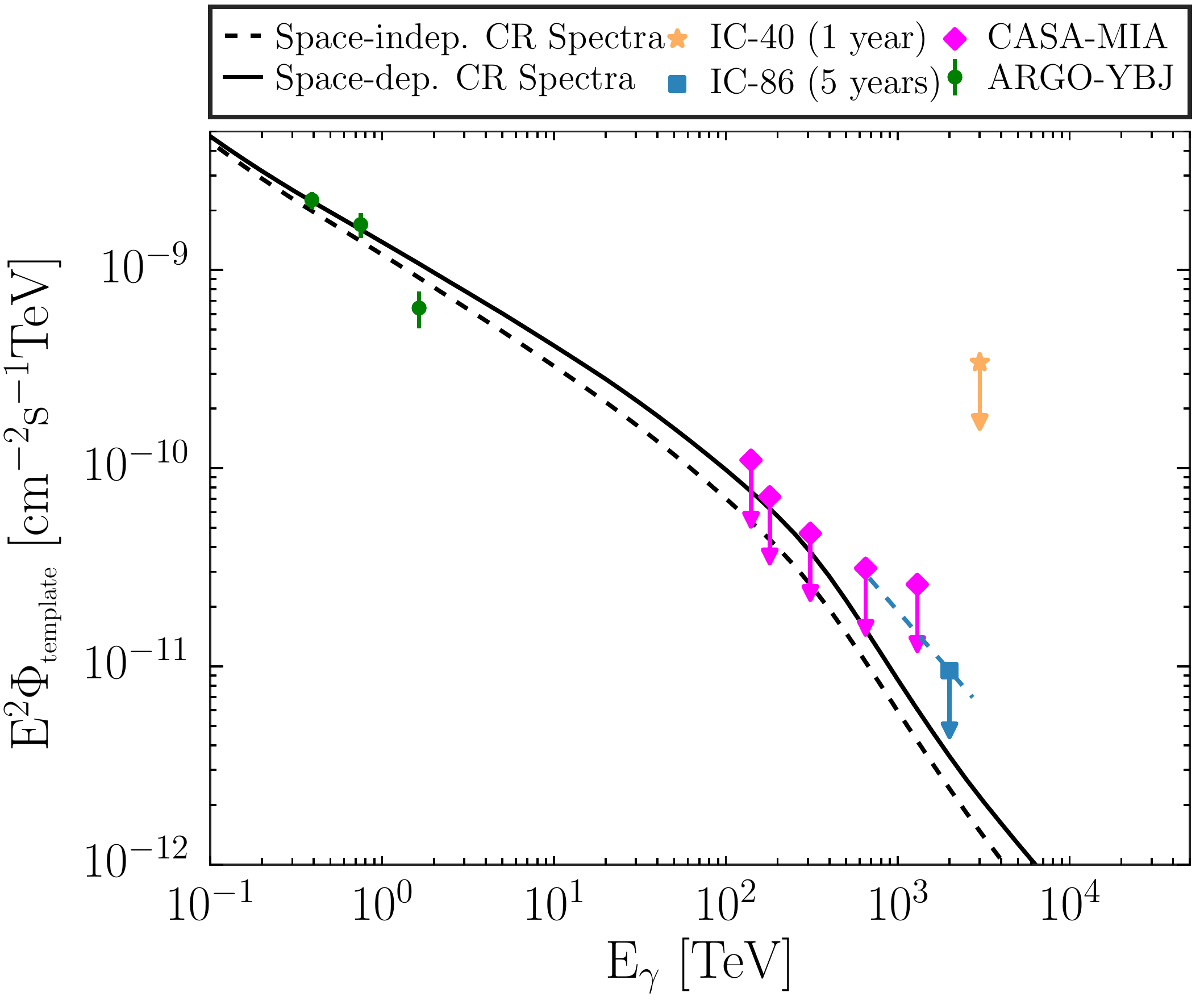}}
  \vspace{-0.3cm}
  \caption{Left: Sensitivity and discovery potential of gamma ray searches in IceCube as a function of declination and assuming two different spectral indices \cite{IceCube3}. Also shown are the H.E.S.S. sources in the field of view, with their fluxes extrapolated to $2\,\mathrm{PeV}$. Right: The IceCube $90\%$ confidence level upper limit on the gamma ray flux from from the galactic plane in IceCube's field of view. The spatial distribution of gamma ray emission is given by the $\pi^0$ decay component of the Fermi-LAT diffuse emission model \cite{Fermi-LAT}. Also shown are results reported by the CASA-MIA \cite{CASA-MIA} and ARGO-YBJ \cite{ARGO-YBJ} collaborations.}
  \label{fig:GRHess}
  \vspace{-0.3cm}
  
\end{figure}

The selection of muon-poor gamma ray showers is based on various observables obtained from IceTop, including individual tank charges, $q_i$, their lateral distances to the reconstructed shower axis, $r_i$, and the times of the recorded hits, $\Delta t_i$. Here $i$ represents each individual IceTop tank. These observables are used to derive three two-dimensional PDFs, mapping the $r$-$\Delta t$-space, the $q$-$r$-space, and the $q$-$\Delta t$-space, as described in Ref.~\cite{IceCube3}. Corresponding likelihood functions are calculated using CORSIKA simulations with Sibyll~2.1 and various spectrum assumptions for a gamma ray hypothesis. The background PDFs are obtained from IceTop data, which is strongly dominated by cosmic ray events. The ratio of these signal and background likelihoods is then obtained individually for each bin of the size $0.1$ in $\log_{10}(S_{125})$ and for zenith angle bins of $0.05$ in $\cos(\theta)$. The sum of the three resulting logarithmic likelihood ratios, as well as the energy proxy $S_{125}$, and the reconstructed zenith angle, are subsequently used as input to a random forest classifier \cite{scikit-learn}. In addition, a simple geometrical factor \cite{IceCube3} is used as input variable to represent the containment of the shower in the deep ice detector. The random forest classifier has been optimized using a cross-validation grid search, and for this analysis, events with a score above $0.7$ are considered to be gamma ray candidates. Since the expected gamma ray flux depends on an assumption of the spectrum, multiple classifiers are used where each year is separately trained. 

This event selection was applied to IceTop data taken between May 2011 and May 2016 \cite{IceCube3}. The resulting dataset is used to test a variety of sources for their gamma ray emission, as discussed in the following. All considered sources in this analysis are tested using an unbinned log-likelihood ratio, as described in detail in Ref.~\cite{Braun1}, that uses the likelihood functions given in Ref.~\cite{IceCube3}. The flux dependence on the spectral index assumed in this analysis reaches its minimum near $\sim 2\,\mathrm{PeV}$. The sensitivity, discovery potential, and flux limits of this analysis are therefore derived at gamma ray energies of $2\,\mathrm{PeV}$. 

The hottest spot in the sky is located at $-73.4^\circ$ in declination and $148.4^\circ$ in right ascension, with a pre-trial p-value of $4\cdot 10^{-5}$ and best-fit spectral index of $\gamma=2.9^{+0.3}_{-0.3}$. The post-trial p-value is $0.18$ and is calculated by comparing a test statistic to the background expectation test statistic. The corresponding sensitivity and discovery potential for point-source searches, assuming two different spectral indices for the gamma ray flux at $2\,\mathrm{PeV}$, is shown in Figure~\ref{fig:GRHess} (left) as a function of declination. There are 15 known TeV gamma ray sources in the field-of-view of this analysis, reported by H.E.S.S. \cite{HESS1}. Since there is no evidence of a cut-off in the reported spectra, they can be extrapolated to $2\,\mathrm{PeV}$. For this extrapolation, the attenuation of gamma rays is taken into account for each source according to its reported distance using the model from Ref.~\cite{Lipari1}. If the distance of a source is unknown, an approximate distance to the galactic center of $8.5\,\mathrm{kpc}$ is assumed. These $15$ H.E.S.S. sources, extrapolated to the expected fluxes at $2\,\mathrm{PeV}$, are also shown in Figure~\ref{fig:GRHess} (left). The most significant individual source, H.E.S.S. J1427-608 \cite{HESS2}, has a pre-trial p-value of $0.07$ and a best-fit spectral index of $\gamma=3.2^{+0.8}_{-0.7}$. The post-trial p-value of $0.65$ is consistent with the background expectation. 

Furthermore, a correlation of PeV gamma rays with high-energy starting track neutrino events in IceCube \cite{IceCube4} was investigated and no correlations were found. A $90\%$ confidence level upper limit of~$1.07 \cdot 10^{-19}\,\mathrm{cm}^{-2}\mathrm{s}^{-1}\mathrm{TeV}^{-1}$ is placed on the gamma ray flux at 2 PeV for a source class consistent with the neutrino directions reported in Ref.~\cite{IceCube4}, assuming an $E^{-2}$ spectrum.

To test for a diffuse flux from the galactic plane, the template likelihood method \cite{Braun1} was employed, with the signal template taken to be the $\pi^0$ decay component of the Fermi-LAT diffuse emission model \cite{Fermi-LAT}. The observed flux is consistent with the background expectation, and a $90\%$ confidence level upper limit of $2.61 \cdot 10^{-19}\,\mathrm{cm}^{-2}\mathrm{s}^{-1}\mathrm{TeV}^{-1}$ is placed on the gamma ray flux from the galactic plane at 2 PeV, assuming an $E^{-3}$ spectrum.


\section{All-Sky Cosmic Ray Anisotropy}
\label{sec:5}

The distribution of cosmic rays observed on Earth appears to be isotropic as a first order approximation. However, various experiments, including IceCube, have previously reported an-isotropies in the arrival direction of cosmic rays, with a complex angular structure that depends on the energy \cite{IceCube5,HAWC2,ARGO-YBJ2}. At large angular scales ($\gtrsim60^\circ$) the anisotropy has an amplitude of about $10^{-3}$, and for small scales roughly $10^{-4}$ with an angular size of $10^\circ$ to $30^\circ$. However, the limited field of view of these observations makes it difficult to interpret these structures in terms of the spherical harmonic components, see for example Ref. \cite{Ahlers1}. In the following, an all-sky measurement of the arrival directions of cosmic rays with median energies of $10\,\mathrm{TeV}$ is discussed. The measurement uses in-ice data taken by IceCube between May 2011 and May 2016 ($\sim 2.8\cdot10^{11}$ events), and data from the HAWC Observatory \cite{HAWC1}, taken from May 2015 to May 2017 ($\sim 2.8\cdot10^{10}$ events) \cite{IceCube6}.

\begin{figure}[b]
\vspace{-0.2cm}

  \centering
  \includegraphics[width=0.48\textwidth]{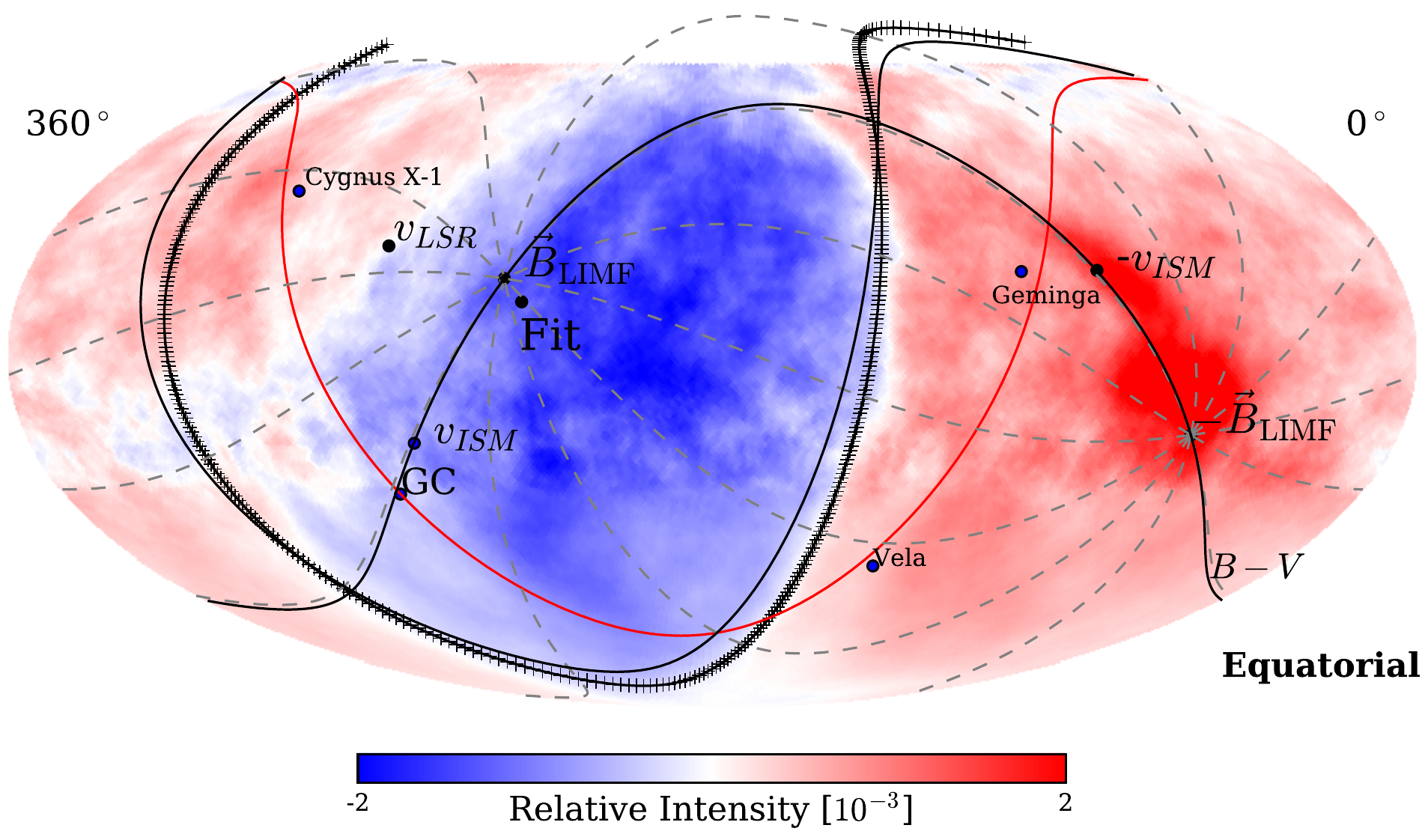}\quad
  \includegraphics[width=0.48\textwidth]{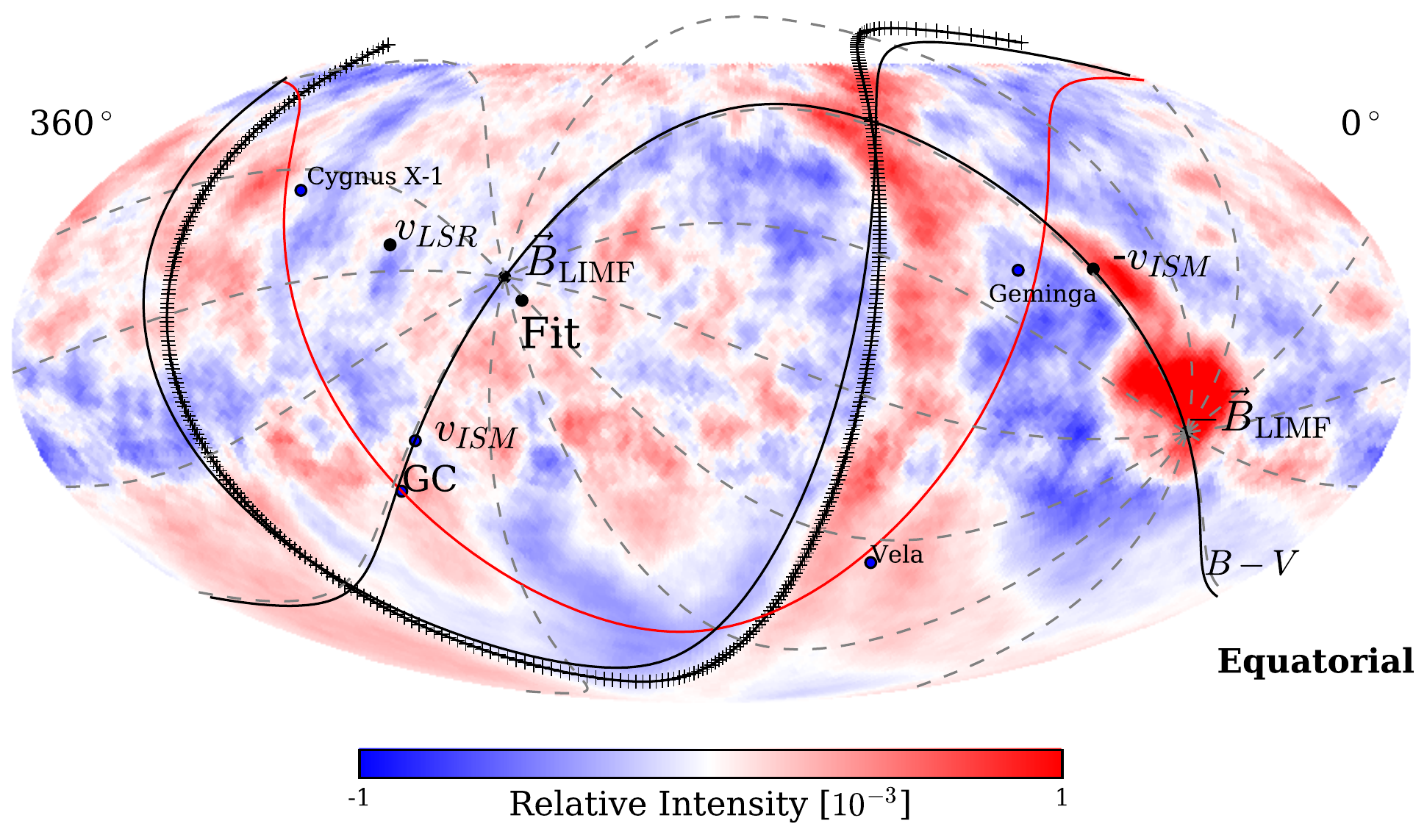}
  \vspace{-0.2cm}
  
  \caption{Mollweide projection sky maps of the relative intensity of cosmic rays at $10\,\mathrm{TeV}$ (left) and after subtracting the multipole fit from the large-scale map (right, see text for details) \cite{IceCube6}. A fit to the boundary between large scale excess and deficit regions is shown as black crossed line and the galactic plane is indicated by the red line. The magnetic equator from Ref.~\cite{Zirnstein}, as well as the $B-V$ plane, are shown as black lines.}
  \label{fig:AnisoRelIntensity}
  \vspace{-0.6cm}
  
\end{figure}

The data collected by the two experiments is processed in order to select cosmic ray events and to improve the reconstruction quality and angular resolution of passing events. A detailed description of the selection criteria can be found in Ref.~\cite{IceCube6}. The cosmic ray energy estimation for IceCube events is based on the number of hit DOMs and the reconstructed arrival direction, as described in Ref.~\cite{IceCube5}. The energy reconstruction in HAWC is discussed in detail in Ref.~\cite{HAWC2}. Additional cuts are applied to select events with energies consistent between both detectors, such that the reconstructed median energy of surviving events is $\sim 10\,\mathrm{TeV}$. The median angular resolution after applying all cuts is about $\sim 3^\circ$ for IceCube events and better than $\sim 0.4^\circ$ for HAWC, respectively. The residual anisotropy is the relative deviation from an isotropic flux expectation, which can be derived for each bin in a sky map. The isotropic expectation is determined based on experimental data in order to account for detector-dependent rate variations by averaging along each declination band, as described in Ref.~\cite{IceCube6}. In order to reconstruct the anisotropy, the maximum log-likelihood method described in Ref.~\cite{Ahlers1} is applied. The corresponding significance is calculated using a generalization of the techniques introduced in Ref.~\cite{LiMa}.

\begin{figure}[t]
\vspace{-0.4 cm}
  \mbox{\hspace{2. cm}
  \includegraphics[width=0.67\textwidth]{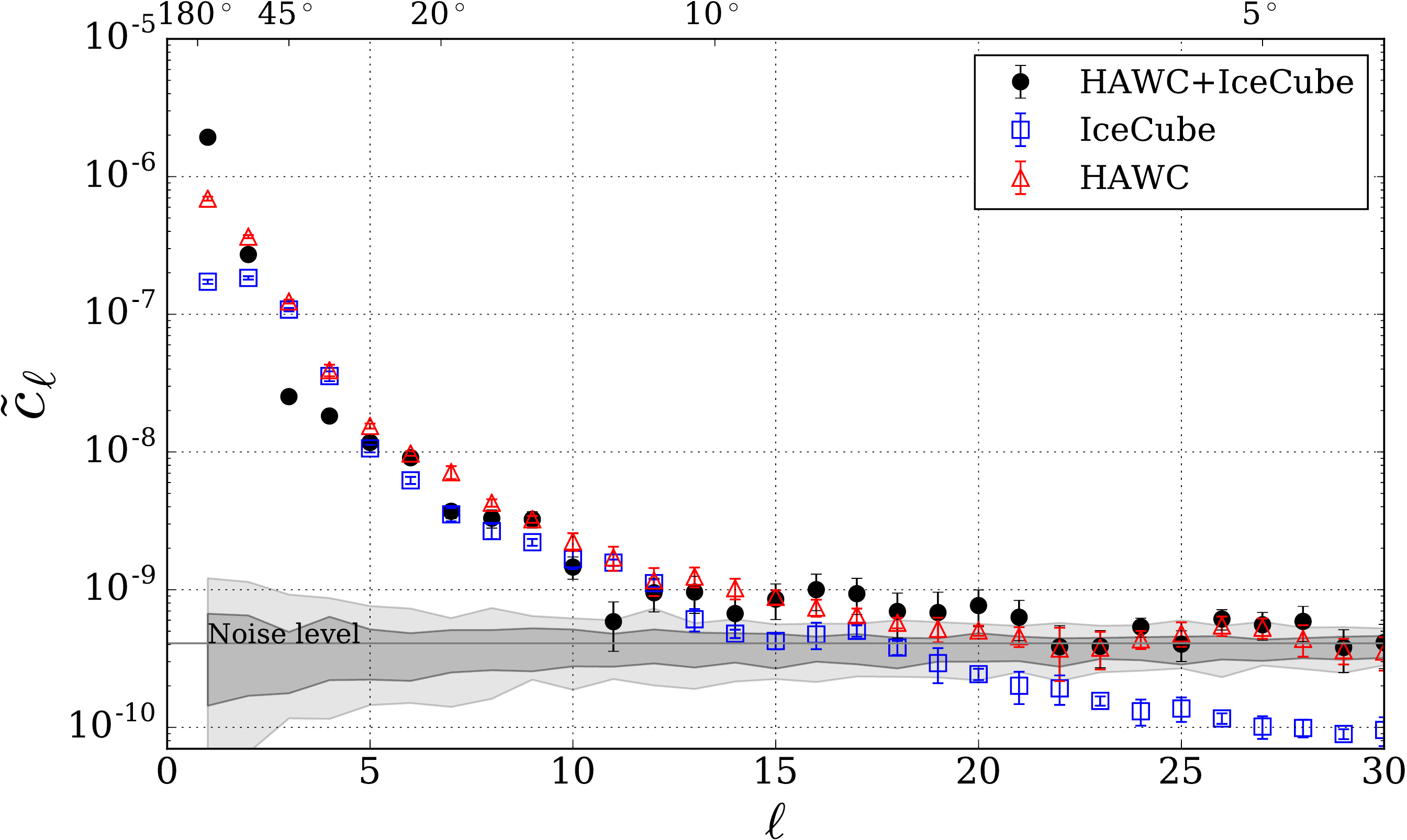}}
  \vspace{-0.3cm}
    \caption{Angular power spectrum of the cosmic ray anisotropy at $10\,\mathrm{TeV}$ \cite{IceCube6}. The grey band represents the $90\%$ confidence level around the level of statistical fluctuations for isotropic sky maps which is dominated by the limited statistics of HAWC data.}
  \label{fig:AnisoPowerSpectrum}
  \vspace{-0.5cm}
  
\end{figure}

The resulting relative intensity map is shown in Figure~\ref{fig:AnisoRelIntensity} (left) where a smoothing procedure was applied using a top-hat function in which each pixel's value is given by the average of all pixels within $5^\circ$. The map is decomposed in spherical harmonic components, where the dipole component, which has most of the power, has an amplitude of $A=(1.17\pm 0.01)\cdot 10^{-3}$ with a phase of $\alpha = (38.4\pm 0.3)^\circ$. The corresponding total systematic uncertainties are $\delta A\simeq 0.006\cdot 10^{-3}$ and $\delta \alpha\simeq 2.6^\circ$. The relative intensity and significance maps after subtracting the fitted multipole from the spherical harmonic expansion with multipole components $\ell \leq 3$ are also shown in Figure~\ref{fig:AnisoRelIntensity} (right). Structures smaller than $\sim 60^\circ$ are observed in this map. The fitted quadrupole component has an amplitude of $A\simeq 6.8 \cdot 10^{-4}$ with a phase of  $\alpha =(20.7 \pm 0.3)^\circ$. These measurements are consistent with those previously reported \cite{IceCube5}. Also shown in Figure~\ref{fig:AnisoRelIntensity} is a fit to the boundary between large-scale excess and deficit regions, as well as the magnetic equator and the plane containing the local interstellar medium magnetic field and velocity, as described in Ref.~\cite{Zirnstein}. 

The angular power spectrum, $\tilde C_\ell$, is derived for the multipole components, $\ell$. It provides an estimate of the significance of structures at different angular scales of $180^\circ/\ell$, and is shown for this analysis in Figure~\ref{fig:AnisoPowerSpectrum}. A discrepancy between the two individual datasets is observed, which is caused by the partial sky coverage in both detectors, and is discussed along with other systematic effects in detail in Ref.~\cite{IceCube6}. This combined analysis provides the first all-sky measurement of the cosmic ray anisotropy at $10\,\mathrm{TeV}$ that removes all biases from the field of view limitations. The resultant angular power spectrum therefore contains valuable information on the cosmic ray distribution and propagation in our galaxy. It enables the determination of the direction of the local interstellar magnetic field, for example, by accounting for the entire angular structure of the anisotropy \cite{Ahlers2}. This measurement, in turn, provides the opportunity to estimate the missing North-South dipole component.


\section{Density of GeV Muons}\label{sec:6}

\begin{wrapfigure}{R}{0.45\textwidth}
\vspace{-1.0 cm}

 \mbox{\hspace{-0.6cm}
  \includegraphics[width=0.53\textwidth]{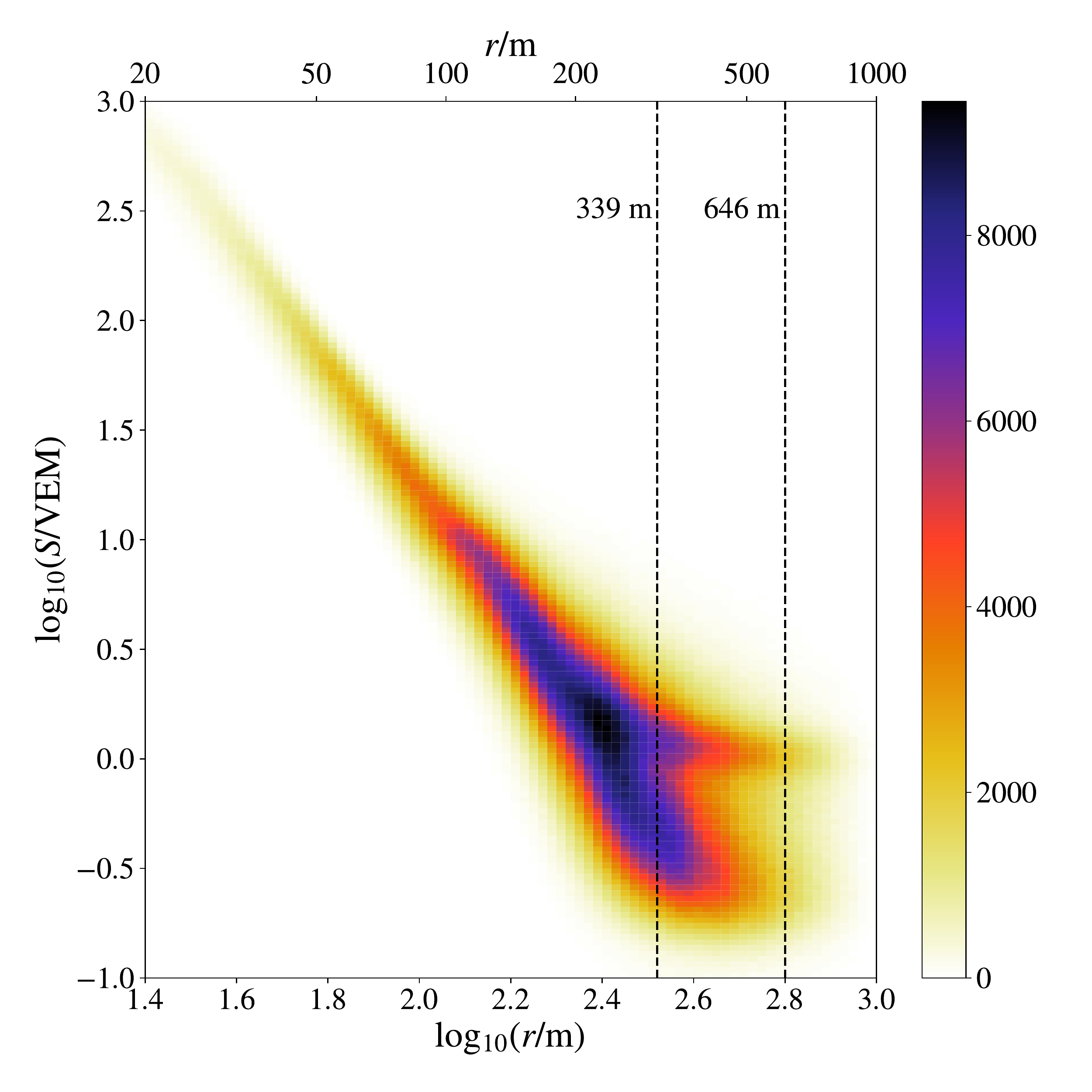}}
  \vspace{-0.8 cm}
  
  \caption{IceTop signals as a function of the lateral distance $r$ of vertical events ($\theta\leq 18^\circ$) with energies between $10\,\mathrm{PeV}$ and $12.5\,\mathrm{PeV}$.}  
  \label{fig:MuonThumb}
\end{wrapfigure}

While the bulk of particles close to the shower axis is highly dominated by electrons, muons become the dominant shower constituents at large distances (several hundred meters) from the core. As previously described in Section~\ref{sec:4}, single tank hits in the shower periphery are predominantly produced by single muons, and therefore need to be considered for muon sensitive analyses. Figure~\ref{fig:MuonThumb} shows the tank signals as a function of the distance to the reconstructed shower axis for shower energies of $10\,\mathrm{PeV}\leq E_0\leq 12.5\,\mathrm{PeV}$ ($\theta\leq 18^\circ$). While most of the signals follow approximately the shape of the LDF from Equation (\ref{eq:S125}), at large distances a structure around $S(r)=1\,\mathrm{VEM}$ becomes visible. This population consists of tanks which measured a single muon and is used to determine the muon content of air showers at large distances.

The GeV muon content of air showers was measured using data taken by IceTop between June 2010 and May 2013 \cite{IceTop6}. The standard IceTop selection is applied \cite{IceTop2} that is also used for the analysis of the cosmic ray spectrum (described in Section~\ref{sec:2}). However, similar to the gamma ray searches, single tank hits are considered for further analysis and hit cleanings similar to those described in Section~\ref{sec:4} are applied. The primary cosmic ray energy is determined using the energy proxy $S_{125}$ and accounting for the snow accumulation, as described in Section~\ref{sec:2}. In addition, only events with zenith angle directions below $\theta\leq 18^\circ$ are considered. After applying all selection criteria, more than $\sim 1.8\cdot 10^7$ events remain in the dataset for further analysis.

\begin{figure}[t]
\vspace{-0.2cm}

  \mbox{\hspace{-0.6cm}
  \includegraphics[width=.51\textwidth]{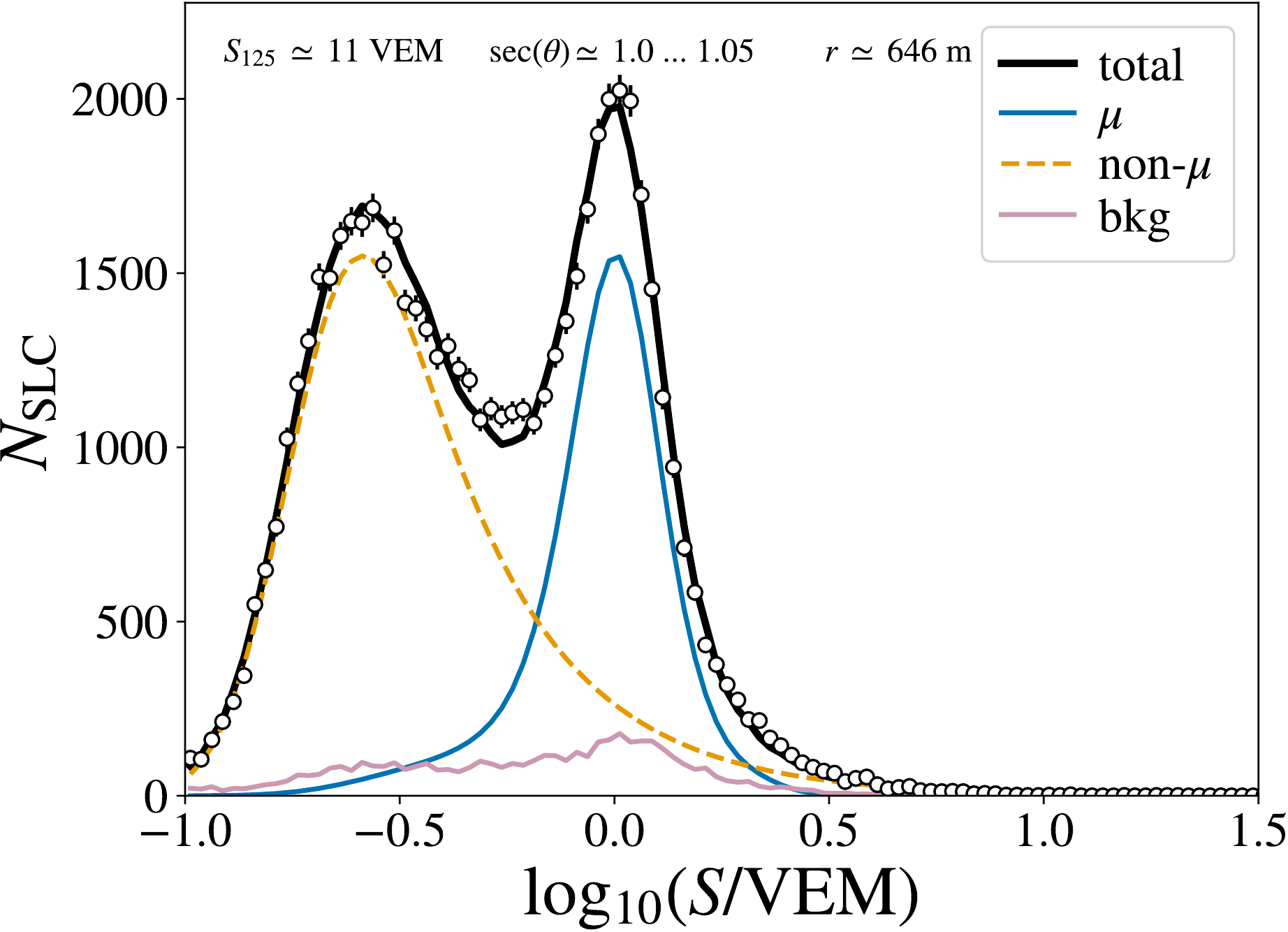}\hspace{-0.0cm}
  \includegraphics[width=.52\textwidth]{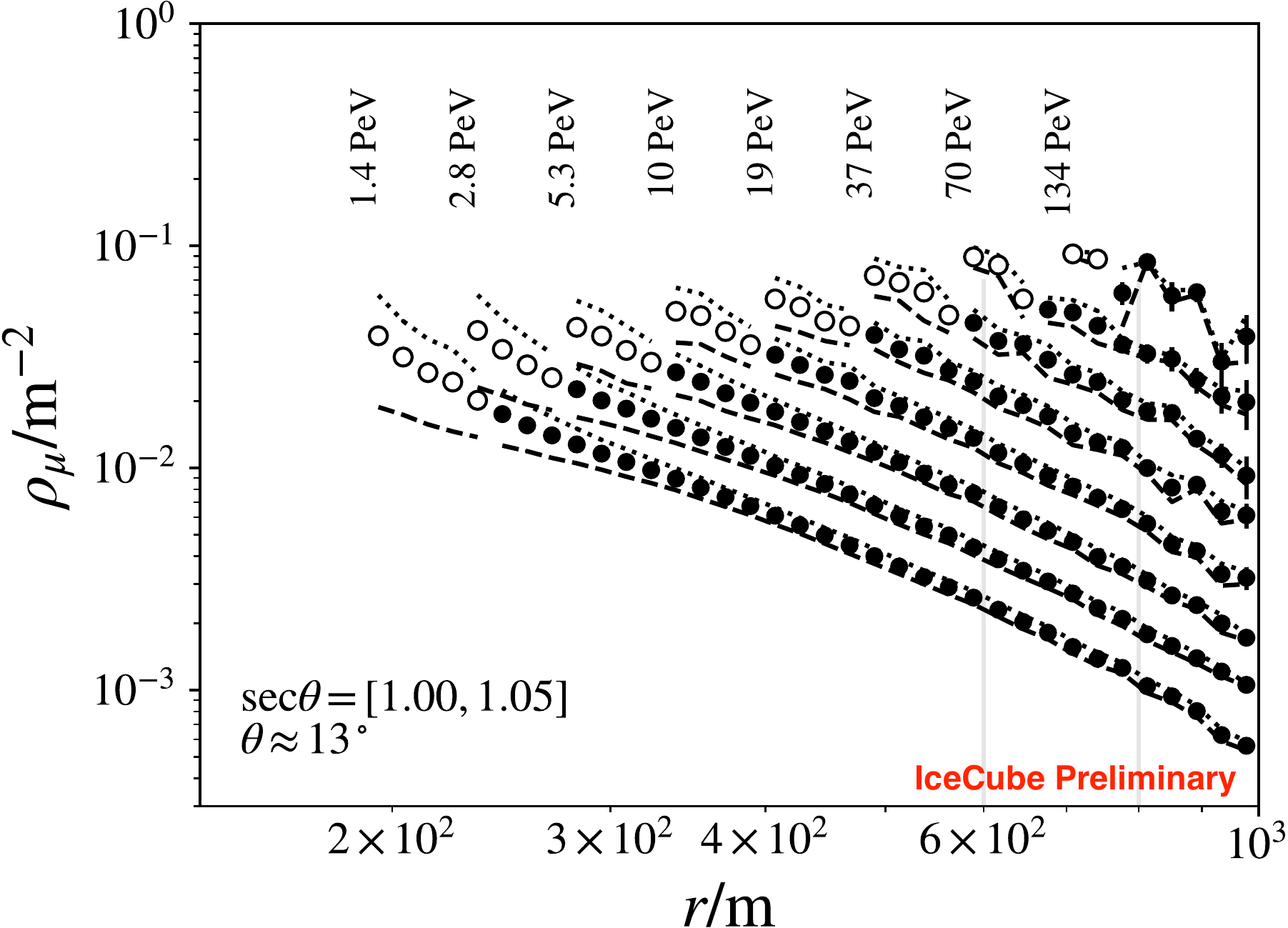}}
  \vspace{-0.6cm}
  
  \caption{Left: Signal distribution at a lateral distance of $646\,\mathrm{m}$ and corresponding fit to the signal model (see text for details). Right: Reconstructed muon density obtained at $600\,\mathrm{m}$ (solid circles) and $800\,\mathrm{m}$ (open squares) as a function of lateral distance for various shower energies \cite{IceTop6}.}
  \label{fig:MuonDensity}
  
  \vspace{-0.2cm}
  
\end{figure}

The number of muons is estimated using a log-likelihood method to fit the signal distributions at a fixed cosmic ray energy, zenith angle direction, and lateral distance. The distributions correspond to vertical slices in Figure~\ref{fig:MuonThumb}, indicated by dashed lines,  and they are fit using a multi-component model, as shown in Figure~\ref{fig:MuonDensity} (left). This semi-analytical model accounts for an electromagnetic component, a muonic component, and for uncorrelated background hits, as described in Ref.~\cite{IceTop6}. The electromagnetic component is described by a simple power law accounting for the threshold behavior of small signals. The muon component is modeled by integer numbers of VEM. The visible track length of muons depends on the initial direction at which they pass through the tank. Inclined muons hitting the central region of the tank deposit charges $\geq 1\,\mathrm{VEM}$, while corner clipping muons deposit less charge \cite{IceTop7}. This smears out the signal distributions and is accounted for in the model description. Also included is the contribution from accidental background hits, which survive the previous cleaning procedures. During this procedure the effects on the signals due to snow accumulation are taken into account. In this way, various free parameters of the model are fit \cite{IceTop6}. Most importantly for this analysis is the mean muon number, $\langle N_\mu\rangle$, which is then divided by the cross-sectional area of the tanks in order to determine the muon density, $\rho_\mu(r)$.

Systematic uncertainties from snow accumulation, the absolute energy scale calibration, and the electromagnetic model in the likelihood fit are considered for this analysis and they are discussed in detail in Ref.~\cite{IceTop7}. However, by construction this analysis contains differences in how the muon density distributions are obtained from simulations and from data. This is caused by the finite resolution of the reconstructed shower core position, direction, and shower energy, as well as by the potential biases due to the event selection and possible effects from the mass composition assumption. It is not possible to disentangle these effects due to the limited statistics in the existing simulated datasets, and a correction therefore needs to be applied to the measured muon densities. This correction is obtained from separate simulations for each year, and systematic uncertainties from this correction, for example due to the hadronic and flux model assumptions, are taken into account in the uncertainty estimates of this analysis \cite{IceTop7}.

\begin{figure}[t]
\vspace{-0.2cm}

  \mbox{\hspace{-0.1 cm}
  \includegraphics[width=0.53\textwidth]{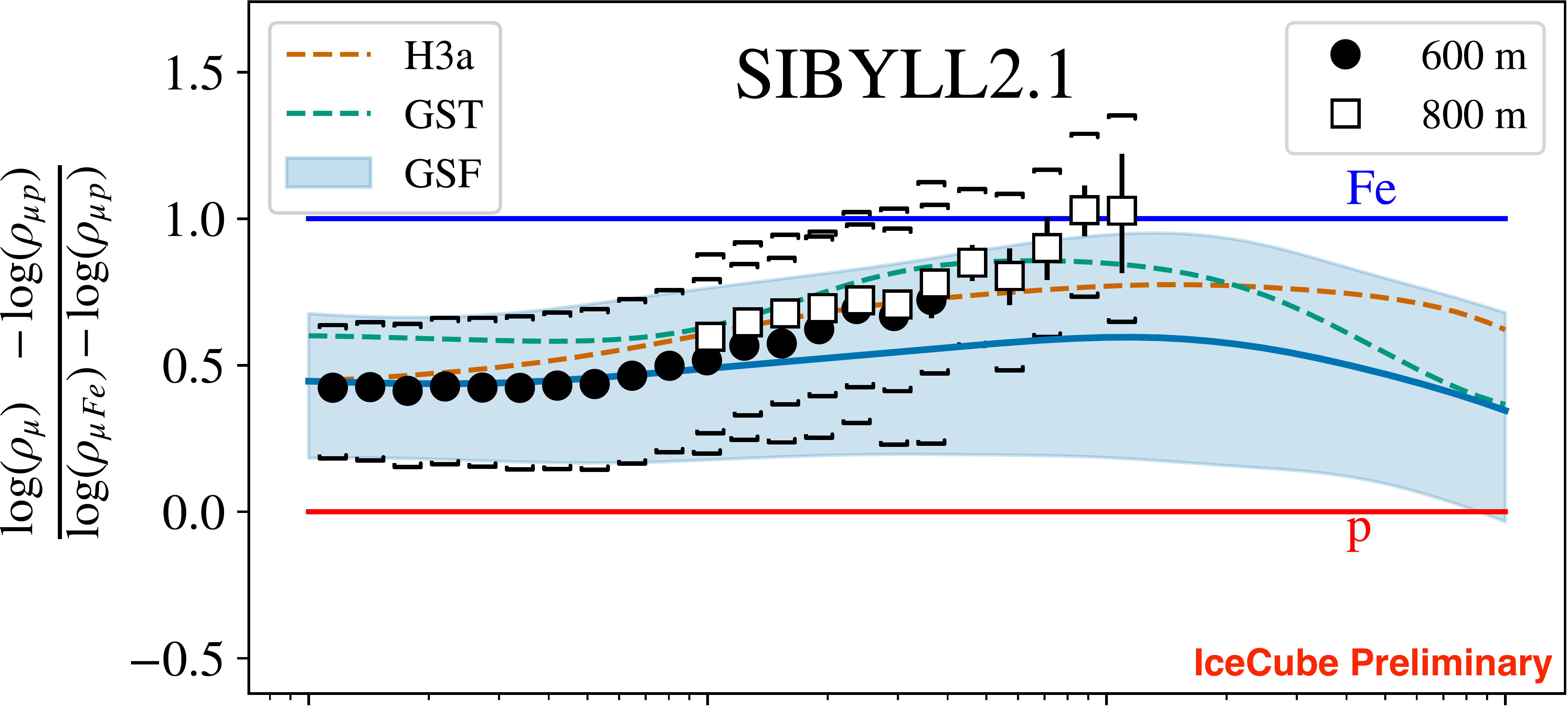}\hspace{-0.06cm}
  \includegraphics[width=0.45\textwidth]{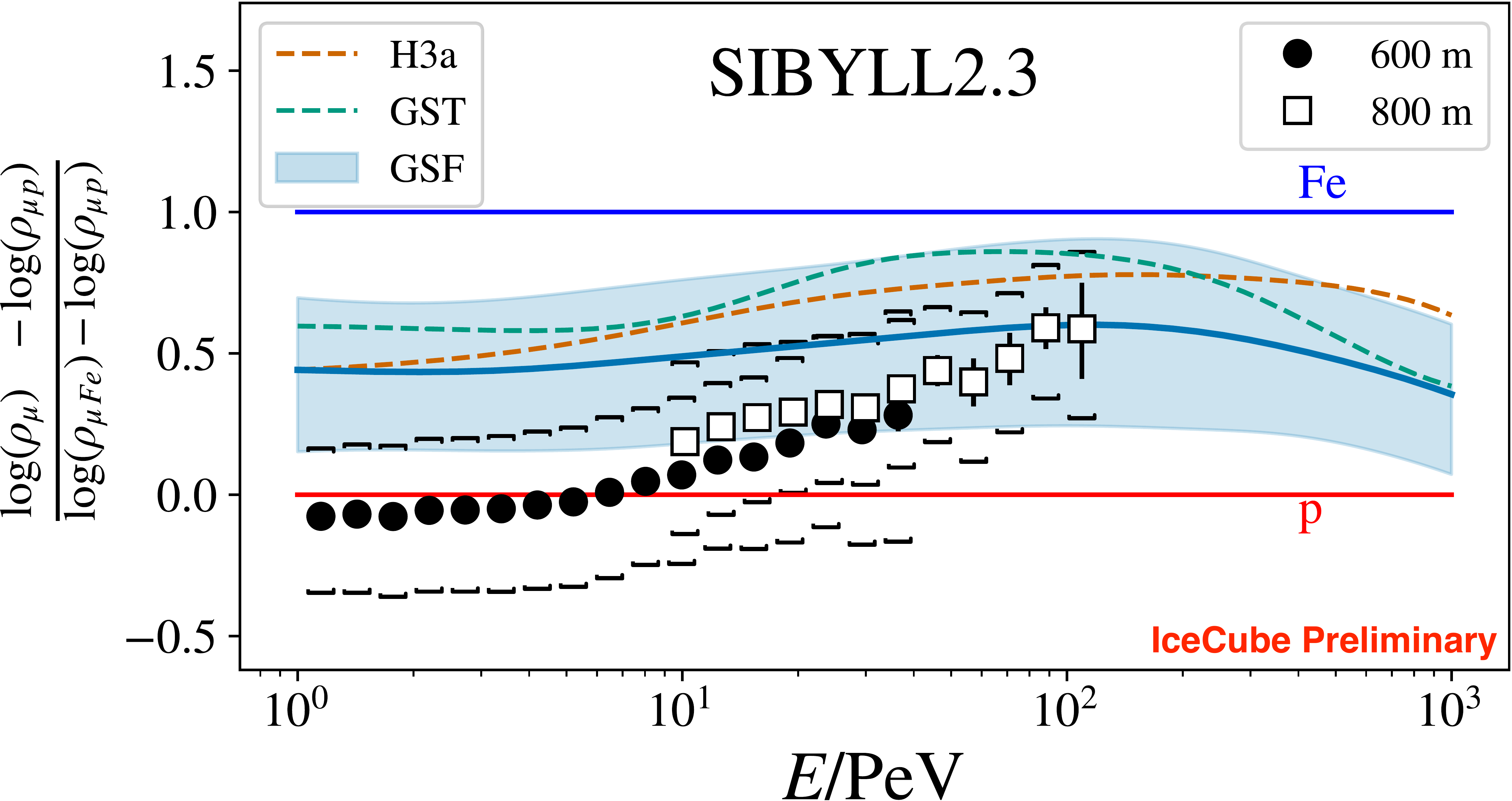}}
  \mbox{\hspace{-0.1 cm}
  \includegraphics[width=0.53\textwidth]{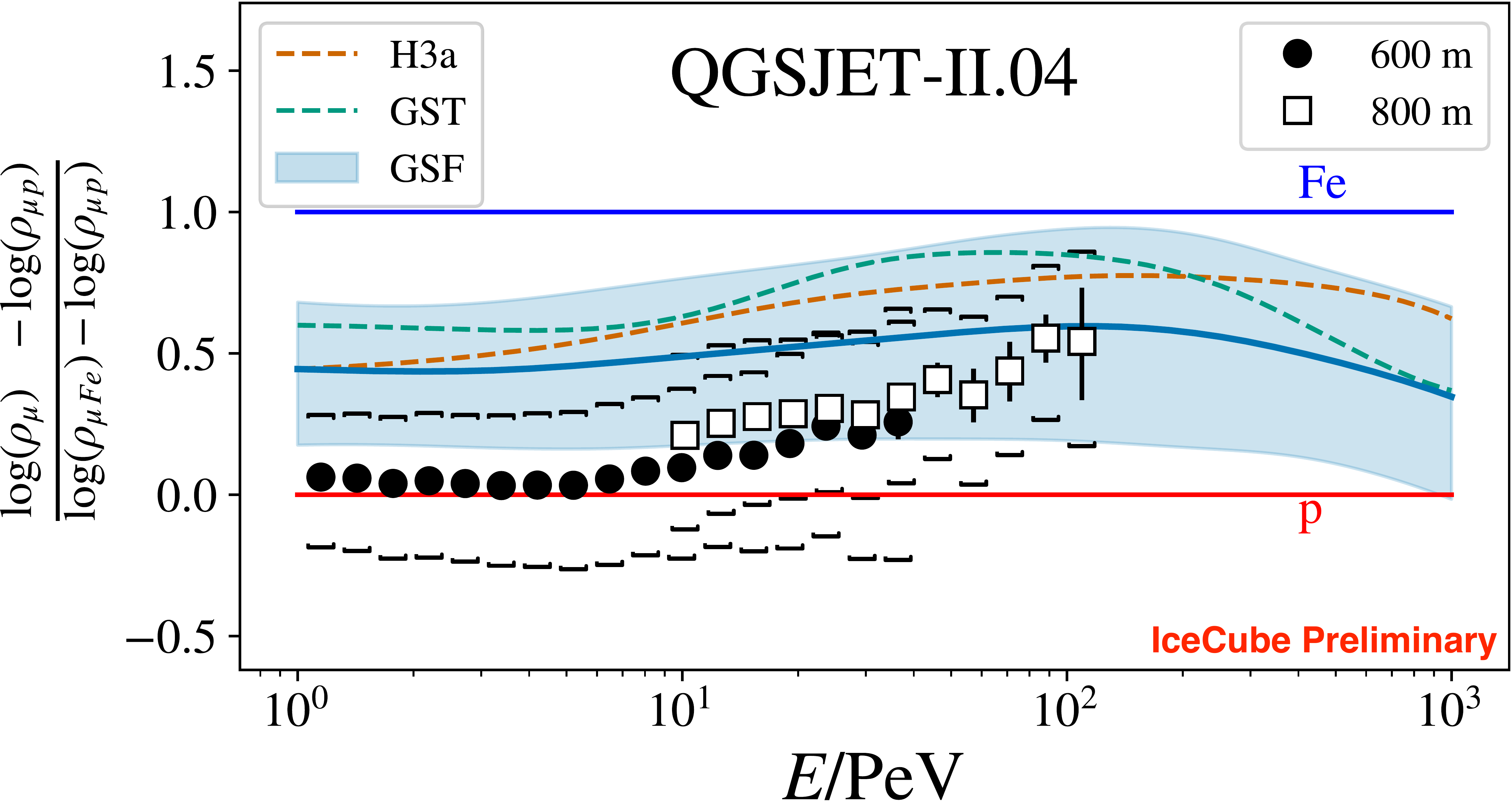}\hspace{-0.06cm}
  \includegraphics[width=0.45\textwidth]{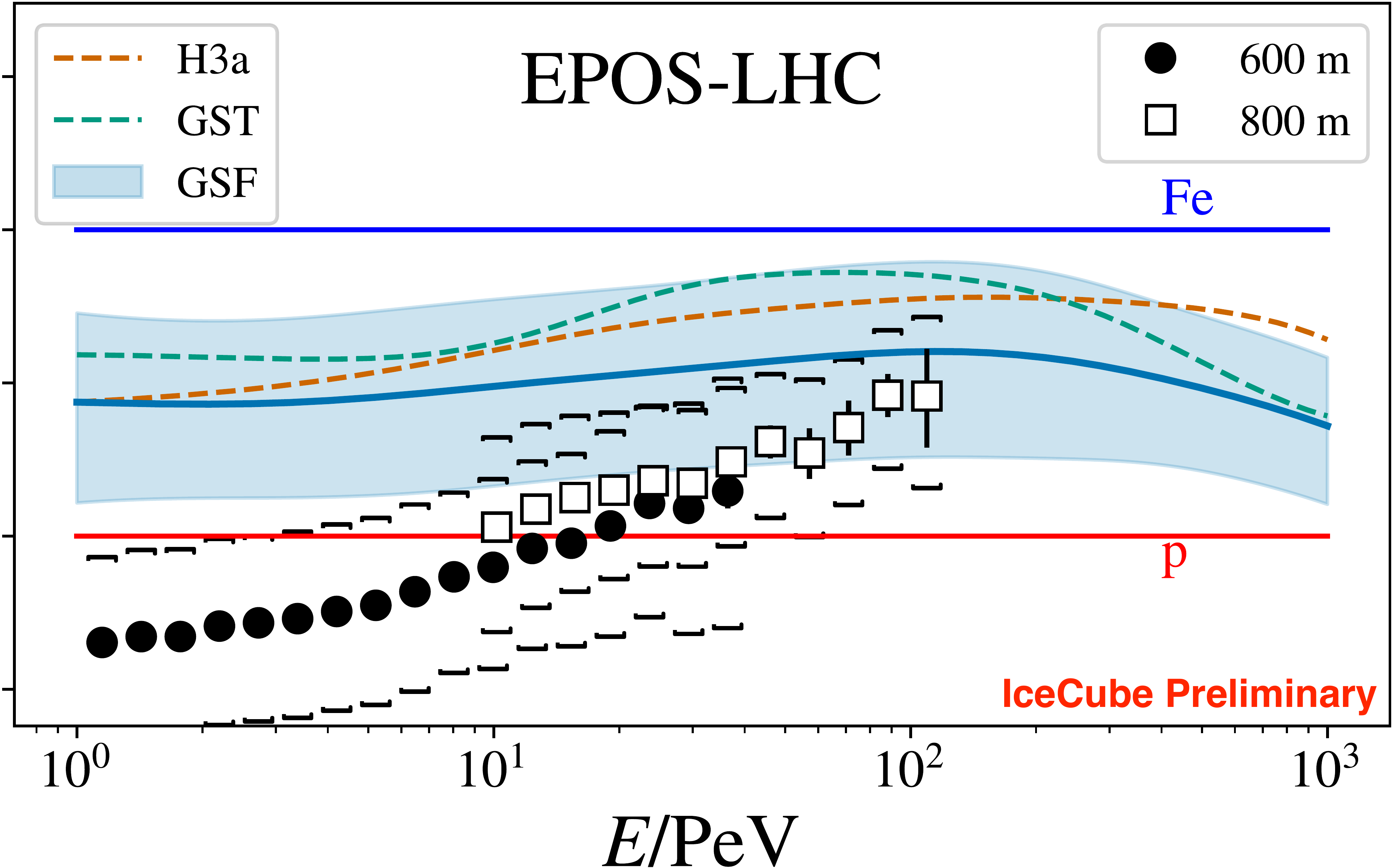}}
  \vspace{-0.2cm}
  
  \caption{Distribution of z-parameters, as defined in the text, as a function of shower energy \cite{IceTop6} compared to various predictions from hadronic models. Also shown are several flux model predictions for comparison. Statistical uncertainties are shown as error bars and brackets indicate systematic uncertainties. Model uncertainties obtained from the GSF flux model are shown as a band.}
  \label{fig:MuonDensityResult}
  \vspace{-0.2cm}
  
\end{figure}

The resulting muon densities as a function of the radial distance from the shower axis are shown in Figure~\ref{fig:MuonDensity} (right) for various cosmic ray energies. Also shown are the corresponding systematic uncertainties. Based on CORSIKA simulations, the mean energy of the muons is estimated to be around $\sim 1\,\mathrm{GeV}$. In further analysis, the muon densities are determined at radial distances of $600\,\mathrm{m}$ and $800\,\mathrm{m}$ as a function of the primary energy. In order to study effects of the hadronic interactions, the measured distributions are compared to simulated data obtained from CORSIKA, using Sibyll~2.1, Sibyll~2.3, QGSJet~II-04, and EPOS-LHC as hadronic models. This is done by introducing a parameter defined as
\begin{equation}
z=\frac{\log(\rho_\mu)-\log(\rho_{\mu,p})}{\log(\rho_{\mu,\mathrm{Fe}})-\log(\rho_{\mu,p})}\, ,
\label{eq:ZFactor}
\end{equation}
where $\rho_\mu$ is the experimentally measured muon density, and $\rho_{\mu,p}$ and $\rho_{\mu,Fe}$ are the muon densities obtained from simulations, assuming a pure proton and pure iron flux, respectively. In this way, this parameter takes the value $z=0$ for proton and $z=1$ for iron primary cosmic rays. The resulting distributions are shown in Figure~\ref{fig:MuonDensityResult} for several hadronic models. The muon densities are obtained at a radial distance of $600\,\mathrm{m}$ for shower energies from $1\,\mathrm{PeV}$ to $40\,\mathrm{PeV}$ (solid circles) and at $800\,\mathrm{m}$ for energies between $9\,\mathrm{PeV}$ and $120\,\mathrm{PeV}$ (open squares). Also shown are the z-parameters obtained from various flux predictions \cite{H3a,GST,GSF}. Predictions using Sibyll~2.3, QGSJet~II-04, and EPOS-LHC as a hadronic model show a larger muon density compared to Sibyll~2.1 simulations. While Sibyll~2.1 and QGSJet~II-04 simulations are bracketed by the pure proton and pure iron predictions, the other two models require a mass composition lighter than protons. A comparison with flux model predictions shows a different change in the mass composition; {\it i.e.} a different slope in the z-parameter, although all models appear to fit the data at different energies. However, recent efforts to study the muon content of air showers in the context of results from various experiments have shown that an energy cross-calibration between the experiments can change the interpretation of experimental data in the context of hadronic models \cite{WHISP1,WHISP2}.

\section{Summary and Outlook}
\label{sec:7}

In this report, the recent results from measurements of cosmic rays with IceCube and IceTop were presented. The resultant cosmic ray energy spectrum from $\sim 250\,\mathrm{TeV}$ up to $\sim 1\,\mathrm{EeV}$, as well as the corresponding mass spectrum above a few PeV, agree with previous measurements within systematic uncertainties. In addition, searches for PeV gamma ray sources were discussed which yield the strongest limits to date on gamma ray fluxes at $2\,\mathrm{PeV}$ for various source candidates considered in this analysis. Moreover, a combined analysis of the arrival directions of cosmic rays at~$\sim 10\,\mathrm{TeV}$, measured with IceCube and HAWC, was presented which enables studying the all-sky anisotropy in this energy range for the first time. Finally, a measurement of the GeV muon content of air showers was discussed in the context of hadronic interaction models. The results can help to improve the modeling of hadronic interactions in air shower simulations in the future. These analyses demonstrate IceCube's unique capabilities to perform a variety of cosmic ray measurements that extend beyond those highlighted in this proceeding.

However, the systematic uncertainties and model dependencies of many cosmic ray measurements can be further reduced in future analyses. Sophisticated calibration devices in the context of an IceCube upgrade \cite{ICRCDawn,IceCubeUpgrade} will help to reduce the systematic uncertainties of composition measurements in IceCube, for example, that are dominated by the light yield uncertainties due to the ice. In addition, potential future surface detector extensions and hybrid detection methods can help to reduce uncertainties of air shower measurements using complementary event information from all detector components \cite{IceTopUpgrade1,IceTopUpgrade2,IceTopUpgrade3} and improved analysis methods \cite{IceTop8,IceTop9}.


\bibliographystyle{unsrt}

\end{document}